\documentclass[12pt]{amsart}
\usepackage{graphics}
\usepackage{amsthm}
\usepackage{amssymb}
\usepackage{amsmath}
\usepackage{amsfonts}
\usepackage{epic}
\usepackage{curves}
\usepackage{epsfig}
\usepackage{bm}
\usepackage{amscd}
\input{psfig.sty}

\newcommand{\baseRing}[1]{\ensuremath{\mathbb{#1}}}
\newcommand{\Z}{\baseRing{Z}}
\newcommand{\C}{\baseRing{C}}
\newcommand{\N}{\baseRing{N}}
\newcommand{\R}{\baseRing{R}}

\theoremstyle{plain}

\theoremstyle{definition}

\def\be{\begin{equation}}
\def\ee{\end{equation}} 
\title{Liouville theory and logarithmic solutions to Knizhnik-Zamolodchikov equation}

\author{Gast\'on Giribet}
\address{Institute for Advanced Study. Einstein Drive, Princeton NJ08540}
\email{gaston@ias.edu}
\thanks{G.G. Institute for Advanced Study. Einstein Drive, Princeton NJ08540, USA}

\author{Claudio Simeone}
\address{Instituto de Astronom\'{\i}a y F\'{\i}sica del Espacio and Departamento de F\'{\i}sica, Universidad de Buenos Aires. Ciudad Universitaria (1428), Buenos Aires, Argentina.}
\email{simeone@iafe.uba.ar}
\thanks{C.S. Instituto de Astronom\'{\i}a y F\'{\i}sica del Espacio and Departamento de F\'{\i}sica, Universidad de Buenos Aires. Ciudad Universitaria (1428), Buenos Aires, Argentina.}

\begin{document}


\begin{abstract}
We study a class of solutions to the $SL(2,\R )_k$ Knizhnik-Zamolodchikov equation.

First, logarithmic solutions which represent four-point correlation functions describing string scattering processes on three-dimensional Anti-de Sitter space are discussed. These solutions satisfy the factorization {\it ansatz} and include logarithmic dependence on the $SL(2,\R )$-isospin variables. Different types of logarithmic singularities arising are classified and the interpretation of these is discussed. The logarithms found here fit into the usual pattern of the structure of four-point function of other examples of $AdS/CFT$ correspondence. Composite states arising in the intermediate channels can be identified as the phenomena responsible for the appearance of such singularities in the four-point correlation functions. In addition, logarithmic solutions which are related to non perturbative (finite $k$) effects are found.

By means of the relation existing between four-point functions in Wess-Zumino-Novikov-Witten model formulated on $SL(2,\R)$ and certain five-point functions in Liouville quantum conformal field theory, we show how the reflection symmetry of Liouville theory induces particular $\Z _2$ symmetry transformations on the WZNW correlators. This observation allows to find relations between different logarithmic solutions. This {\it Liouville description} also provides a natural explanation for the appearance of the logarithmic singularities in terms of the operator product expansion between degenerate and puncture fields. 
\end{abstract}



\maketitle




\section{Introduction}

The analysis of four-point correlation functions in the world-sheet formulation of string theory on $AdS_3$ is an important example within the context of the study of non-compact conformal field theory. The reason for this is because this particular case presents several subtle aspects which are characteristic of this class of non trivial CFT's; for instance, this theory admits a non unique fusion matrix\footnote{The monodromy of this theory was also studied in references \cite{rass,petko}} \cite{ponsot}, and this feature is related with another particular aspect, {\it i.e.} the existence of an additional (a fourth) singular point in the world-sheet \cite{ponsot,petko,mo3,teschner} which is located in the middle of the moduli space \cite{mo2}. This CFT has non trivial factorization properties since additional restrictions are required in order to render the operator product expansion well defined \cite{mo3}; this peculiarity is associated to the requirement of additional constraints for the unitarity of the spectrum and for locality in the dual conformal field theory ({\it i.e.} dual in the sense of $AdS/CFT$ correspondence \cite{mo3}).

Then, a detailed study of the analytic structure of these correlators turns out to be an interesting topic of investigation, as in the context of the question about the consistency of string theory formulated on curved space-times as also in the context of the study of formal aspects of two-dimensional conformal field theories. The principal study of two and three-point functions in WZNW model formulated on $SL(2,\C )/SU(2)$ was carried out in references \cite{teschner}.

In this paper we study the logarithmic singularities arising in the solutions of the Knizhnik-Zamolodchikov (KZ) equation \cite{kz}. 

This work presents a survey of solutions to the KZ equations by using two different techniques. First, we analyse the factorization {\it ansatz} in order to address the study of logarithmic singularities arising in these four-point functions representing tree-level scattering processes of string theory in $AdS_3$. Secondly, we make use of the correspondence existing between the conformal model on $SL(2,\R )$ and Liouville field theory in order to explore several features of the solutions and the structure of four-point functions; in particular, we study the role of the spectral flow symmetry of the $\hat {sl(2)}_k$ algebra which turns out to be a crucial point in the framework. We show, in that manner, a natural explanation for the appearance of logarithmic singularities in the four-point correlators; this enables us to present a {\it dual Liouville picture} for the $AdS_3$ processes.

\subsection{On logarithmic solutions}

Logarithmic solutions of KZ equation have been previously studied in the literature in different contexts \cite{log} (see the seminal work \cite{gurarie}). Here, we focus our attention on the solutions representing physical processes interpreted as string interactions in $AdS$ space-time.  The appearance of logarithms in the explicit computation of several examples of four-point processes in $AdS$ was a confusing issue during the first stages of $AdS/CFT$ correspondence; and it was finally understood and extensively explained in the context of supergravity description by the authors of references \cite{liu,ras1,ras2,ras3,ras4}.

These previous works motivate the question about whether these features are also present in the three dimensional case. The particular interest of the three dimensional case is based on two points: first, the boundary conformal field theory associated to string theory on $AdS_3$ presents special features due to its non-compactness and, thus, the study of its properties turns out to be an interesting subject of investigation by itself. On the other hand, because string theory on $AdS_3$ in NS-NS background can be described in terms of Wess-Zumino-Novikov-Witten model, it is an interesting opportunity to study these features in an exact conformal example beyond the field theory limit.

Then, we dedicate the second section of this work to the study of the appearance of logarithmic singularities in the expression of four-point correlators described by KZ equation. We show that the analytic structure in this case holds in the expected pattern of the $AdS/CFT$ correspondence. In addition, we will be able to discuss non-perturbative effects which, consequently, could not be investigated in the field theory computation.

In \cite{mo3} it was commented that logarithms stand in the solutions of $\hat {sl(2)} _k$ KZ equation for particular configurations of the interacting states. In that work it was assumed that the momenta of the states involved in the interaction processes were generic enough so that no logarithmic terms appear in correlators; it is our intention here to explicitly work out the details of such {\it special} cases. Also in \cite{mo3} it was discussed how two particle states appear in the intermediate channels of the factorization limit of four-point functions in $AdS_3$. We will show here that the interpretation of logarithmic singularities as interaction processes involving two-particle states is consistent. In fact, we will be able to show that the conditions to be satisfied between the incoming and outgoing momenta of states when logarithms appear in KZ equation are similar to those which have been found in the analysis of the other examples of $AdS/CFT$ correspondence and correspond to certain tuning of quantum numbers of the four interacting states.

\subsection{Outline}

In the next section we find solutions to $\hat {sl(2)}_k$ KZ equation which satisfy the factorization {\it ansantz} and contain logarithmic terms in the $SL(2,\R )$-isospin variables. We discuss the interpretation of these in the context of $AdS/CFT$ correspondence (see (\ref{poi}) below).

As a corollary of section 2, we will find how the analysis of the conditions for the appearance of logarithmic solutions teaches us that, in order to understand the structure of four-point function, it would be necessary to previously understand some aspects of the spectral flow symmetry acting on the spectrum of the theory. 

With this purpose, considered as a requirement, we move in the third section to the analysis of a particular representation of four-point functions by using the known relation existing between such observables and similar quantities in Liouville conformal field theory. We dedicate the appendix A to present a detailed discussion about this relation and its integral representation.

Concisely speaking, the two main results are the following:

First, one of the main goals here is to understand how the appearance of logarithmic solutions of the KZ equation is seen in the Liouville field description of the WZNW correlators. Thus, we will be able to show how to reobtain the logarithmic solutions in a rather different way which provides us a dual-Liouville picture of these effects in $AdS_3$.  This Liouville picture leads to a natural explanation for the arising of the logarithmic singularities in terms of the operator product expansion between degenerate and puncture fields (see (\ref{ert67})). 

Secondly, the Liouville description of WZNW correlation functions also provides a useful tool to study non trivial symmetries of KZ equation. By using this, we derive in the appendix B several relations for solutions of the KZ equation which are viewed on the {\it Liouville side} as rather simple $\Z _2$ transformations. {\it e.g.} some of these symmetries, which have been previously signaled in the literature as non trivial properties of $SL(2,\R ) _k$ correlators, appear here as simple identities derived by the reflection properties of Liouville vertex operators (see (\ref{fuba}) and (\ref{n13}) below).

\subsection{String theory on $AdS_3$ from conformal field theory}

Interpreted as a non-linear $\sigma $-model action, the Wess-Zumino-Novikov-Witten model (WZNW) formulated on $SL(2,\R)$ describes the string dynamics on Lorentzian three-dimensional Anti-de Sitter spacetime ($AdS_3$); and in analogous way, its euclidean version is represented by the model formulated on the homogeneous space $SL(2,\C)/SU(2)$.

From the point of view of this geometrical interpretation, the {\it integrated} correlation functions of both models mentioned, which represent string scattering amplitudes, are related by a Wick rotation in the target space. However, some subtleties of this connection remain yet unclear. Indeed, a detailed worldsheet description of string theory formulated on the lorentzian Anti-de Sitter$_3$ space is not yet available, unlike the case of its euclidean version. 

\subsubsection{The spectrum}

The range of quantum numbers characterizing the states in both models requires to be carefully treated in terms of analytic continuation in the space of the representation indices \cite{mo3,teschner,mo1}. For instance, the $SL(2,\R)$ discrete series appear as pole conditions of the analytic continuation of the expression for two-point functions on $SL(2,\C)/SU(2)$. Consequently, an analytic extension of the functional form of the vertex operators is required in order to incorporate the operators which create states belonging to the discrete representations. 


These two-point functions, representing the reflection coefficient $B(j)$, and the three-point functions, representing the structure constants $C(j_1,j_2,j_3)$, were explicitly calculated in \cite{teschner}\footnote{Here we adopt the nomenclature $j$, as usual, to refer to the index which labels the $SL(2,\R )$ representations following the notation introduced in \cite{mo1}; $j_{\mu }$ represents the momentum of the $\mu ^{th}$ string state in the physical application. }. The pole structure of the {\it integrated} correlation functions was studied with detail in \cite{kutasov,mo3} in the context of string theory.

Interesting observations about the physical interpretations of the analytic structure of more general $N$-point correlation funcions in $SL(2,\C)/SU(2)$ were made in \cite{mo3}. In that paper, the authors identified the correlators that do have a clear physical interpretation as string scattering processes in $AdS_3$, showing that for this purpose it is necessary to impose the restriction $\sum _{\mu =1}^N j_{\mu } <k+N-3$ on $N$-point functions. This constraint implies, for instance, that the holographic image of the scattering processes in the boundary conformal field theory are local correlators.

The spectrum of string theory on $AdS_3$ was constructed by Maldacena and Ooguri in \cite{mo1} in terms of discrete ${\mathcal D}_j$ and continuous ${\mathcal C}_j$ representations of $SL(2,\R)_k$ and its extensions induced by the inclusion of the spectral flow symmetry in the framework, which permits to describe the winding sectors by extending the Hilbert space as\footnote{For parameters $\omega \in \Z$, the spectral flow mapping is an automorphism of the $\hat {sl(2)}_k $ algebra and generates new $\omega$-sectors in the spectrum which can be interpreted as $\omega $inding strings in $AdS_3$ despite the {\it trivial} topology of the space-time, which enriched by the existence of the NS-NS field.} ${\mathcal D}_j \oplus {\mathcal C}_j \rightarrow \bigoplus _{\omega } {\mathcal D}^{\omega }_j \oplus {\mathcal C}^{\omega }_j$. The states belonging to the continuous representations are characterized by quantum numbers $j \in \frac 12 + i\R$ while the states of the discrete representations present real values of $j$ restricted by
\be
1 <2j<k-1 \label{izabel}
\ee
in order to render the spectrum of the free theory unitary. This is a sufficient and non-necessary condition which is not guaranteed by Virasoro constraint.

\subsubsection{Vertex operators}

Then, we have to consider the basic objects of the construction, the vertex operators $\Phi _j$, which create states from the $SL(2,\R)$-invariant vacuum and can be associated to integrable functions on the $H^+ _3$ hyperbolic hyperplane, and which we denote here by $\Phi _{j}(z,\bar z;x, \bar x)$. These are primary fields of the Virasoro algebra with conformal dimension
\begin{eqnarray}
h_j=-\frac {j(j-1)}{k-2}
\end{eqnarray}
and form a representation of the $\hat {sl(2)}_k \otimes \hat {sl(2)} _k$ Kac-Moody algebra generated by currents $J^+$, $J^-$, $J^3$ and the corresponding anti-holomorphic copies. The auxiliary variables $(x,\bar x)$ are introduced in the realization of $\Phi $ in order to classify the mentioned $SL(2,\R )$ representations, and induce a realization of the $sl(2)$ algebra in terms of differential operators, namely
\begin{eqnarray*}
J^{a }(z)\Phi _{j}(w,\bar{w};x,\bar{x}) &=&-\frac{1}{\left( z-w\right) } {\mathcal J}^{a } \Phi _{j}(w,\bar{w};x,\bar{x})+...
\end{eqnarray*}
being $a ={+,-,3}$ and
\begin{eqnarray*}
{\mathcal J}^{+}=x^{2}\frac{\partial }{\partial x}+2jx \ \ \ {\mathcal J}^{3}=x\frac{\partial }{\partial x}+j \ \ \ {\mathcal J}^{-}=-\frac{\partial }{\partial x}
\end{eqnarray*}
This operator product expansion, and its corresponding anti-holomorphic part, encodes the whole algebraic structure.

Within the context of the {\it stringy} application of the conformal model, these coordinates ($x,\bar x$) are interpreted as the coordinates for the boundary of $AdS$ space, where the dual conformal field theory (BCFT) is formulated according to the statement of the $AdS/CFT$ correspondence. In these terms, $\Phi (z;x)$ represent the bulk-boundary propagators, whose physical meaning was earlier clarified in \cite{ooguri}. The world-sheet theory coordinates will be denoted by $(z,\bar z)$ as usual. 

The $SL(2,\C)/SU(2)$ homogeneous space can be parametrized in terms of the Gauss decomposition, which leads to obtain the natural metric in Poincare coordinates covering a half of the euclidean $AdS_3$.

The maps between the target space coordinates and the $(x,\bar x)$ variables\footnote{often refered as $SL(2,\R )-${\it isospin coordinates} or {\it Kac-Moody part.}}, which are usually denoted as $\gamma (z) \rightarrow x$, turns out to be of crucial importance in relation with the geometrical interpretation of the bulk-boundary correspondence. These maps provide a clear picture of certain divergences arising in correlation functions \cite{mo3,ooguri}.

Let us introduce some notation: We will denote by $\sigma $ and $\tilde \sigma $ the Weyl transformations defined as follows 
\begin{eqnarray*}
\sigma : \Phi _j \rightarrow \Phi _{\sigma (j)} &=& \Phi _{1-j} \ , \ \ \ \ \ \ \ \ \tilde \sigma : \Phi _j \rightarrow \Phi _{\tilde \sigma (j)} = \Phi _{k-1-j}
\end{eqnarray*}
These transformations leave the Casimir $h_j$ and the {\it reduced} operator $\tilde h_j=h_j-j$ invariant respectively. We will also find convenient to define the $s$-automorphism as
\begin{eqnarray}
s:\Phi _j \rightarrow \Phi _{s(j)} = \Phi _{\frac k2 -j}
\end{eqnarray}
and the \Z$_2$ $r$-transformation, defined in the following way for the case when a set of four indices $j_{\mu }$ is considered, namely
\begin{eqnarray}
r:\Phi _j \rightarrow \Phi _{r(j_{\nu} )} = \Phi _{\frac 12 \sum _{\mu =1}^{4} j_{\mu} - j_{\nu +2}}
\end{eqnarray}
for the set $\{j_1,j_2,j_3,j_4\}$, where the identification $j_{\nu }=j_{\nu '}$ holds if $\nu =\nu '$ Mod $4$. All of these transformations satisfy $\sigma ^2 = \tilde \sigma ^2 = s^2 =r^2= \baseRing{I}$, $\sigma s = s \tilde {\sigma }$ and clearly $\frac 12 \{\tilde \sigma,\sigma \} = \frac 12 \{\sigma , s \} = \frac 12 \{\tilde \sigma, s \} =  \baseRing{I}$. Here, we have used the vectors $\Phi $ to present the definition of the symmetry transformations, but these can be also defined as acting on higher rank tensors, {\it e.g.} ${\mathcal F}_{j_1,\sigma (j_2),j_3,...} = {\mathcal F}_{j_1,1-j_2,j_3,...}$.

Notice that $s$-transformation is closed among the physical spectrum of states of discrete representations (\ref{izabel})  while $\sigma $ and $\tilde \sigma $ exclude the redundant regions $(0,\frac 12)$ and $(\frac {k-1}{2}, \frac k2)$ respectively. We could also have referred to the index $s(j)$ by simply using the natural notation $\tilde j = \frac k2 -j$. On the other hand, $s$-transformation describes a symmetry existing between certain states of the Hilbert space and due to the spectral flow automorphism. Indeed, the $\omega = \pm 1$ spectral flow transformation\footnote{The spectral flow transformations are defined on the modes $J^a_n$ of the currents $J^a(z)=\sum _{n\in \Z} J^a_n z^{-1-n}$ by the following application $J^{\pm}_n \rightarrow J^{\pm}_{n\pm \omega}$ and $J^{3}_n \rightarrow J^{3}_{n} - \frac k2 \delta _{n,0} \omega$. The $s$-transformation maps Kac-Moody primary states into {\it flowed} primary states by commuting the highest and lowest-weight discrete representations.} do not generate new $\hat {sl(2)}_k$ representations, but simply permute highest and lowest-weight representations by relating certain pair of vectors of both sectors $\omega =0$ and $\omega =\pm1$ by means of the identification between index $j$ and index $s(j)$. This observation will be useful for further discussions because some of the conditions for the arising of logarithmic singularities are related by this transformation.

\section{Knizhnik-Zamolodchikov equation and factorization ansatz}

In the first two subsections we review the factorization {\it ansatz} as a proposal to solve the KZ equation. This was built up in references \cite{ponsot,mo3,teschner}.

\subsection{The factorization {\bf {\it ansatz}}}

The four-point correlation functions on the zero-genus topology are determined by conformal invariance \cite{fz4p} up to a factor ${\mathcal F}$ which is a function of the cross ratio $z$ and the indices $\{ j_i,x_i,\bar x_i \}$ which label the representations; namely
\[
{\mathcal A}^{WZNW}_{j_1,j_2,j_3,j_4} = \left\langle \prod _{i=1}^{4} \Phi _{j_{i}}(z_{i};x_{i})\right\rangle
=\prod_{a<b}\left| x_{a}-x_{b}\right| ^{2J_{ab}}\prod_{a<b}\left|
z_{a}-z_{b}\right| ^{2h_{ab}}\left| {\mathcal F}_{j_1,j_2,j_3,j_4}(x,z)\right| ^{2}
\]
being
\[
x=\frac{(x_{1}-x_{2})(x_{3}-x_{4})}{(x_{4}-x_{2})(x_{3}-x_{1})},\qquad z=%
\frac{(z_{1}-z_{2})(z_{3}-z_{4})}{(z_{4}-z_{2})(z_{3}-z_{1})}
\]
and where $h_{34} =h_{2}+h_{1}-h_{3}-h_{4},$ $h_{24}=-2h_{2},$ $h_{14} =h_{2}-h_{1}+h_{3}-h_{4},$ $h_{13}=h_{4}-h_{1}-h_{2}-h_{3};$ and $J_{34} =j_{1}+j_{2}+j_{3}-j_{4},$ $J_{24}=-2j_{2},$ $J_{14} =-j_{1}+j_{2}+j_{3}-j_{4},$ $J_{13}=-j_{1}-j_{2}-j_{3}+j_{4}$. 

The four-point function will be given by certain linear combination of solutions to the KZ equation; {\it i.e.} that which is monodromy invariant.

The KZ equation in the case of WZNW model formulated on $SL(2,\R )$ is the following
\[
(k-2)z(z-1)\frac{\partial }{\partial z}{\mathcal F}_{j_1,j_2,j_3,j_4}(x,z)=\left( (z-1){\mathcal D}_{1}+z%
{\mathcal D}_{0}\right) {\mathcal F}_{j_1,j_2,j_3,j_4} (x,z)
\]
where the differential operators are
\begin{eqnarray*}
{\mathcal D}_1 &=&x^{2}(x-1)\frac{\partial ^{2}}{\partial x^{2}}-\left(
(j_{4}-j_{3}-j_{2}-j_{1}-1)x^{2}+2j_{1}x+2j_{2}x(1-x)\right) \frac{\partial 
}{\partial x}+ \\
&&+2(j_{1}+j_{2}+j_{3}-j_{4})j_{2}x-2j_{1}j_{2} \\
{\mathcal D}_0 &=&-(1-x)^{2}x\frac{\partial ^{2}}{\partial x^{2}}+\left(
(-j_{1}-j_{2}-j_{3}+j_{4}+1)(1-x)-2j_{3}-2j_{2}x\right) (x-1)\frac{\partial 
}{\partial x}+ \\
&&+2(j_{1}+j_{2}+j_{3}-j_{4})j_{2}(1-x)-2j_{2}j_{3}
\end{eqnarray*}
Notice that ${\mathcal D}_0$ and ${\mathcal D}_1$ are related under the interchange $x \leftrightarrow 1-x$ and $j_1 \leftrightarrow j_3$ which is clear from the definition.

Let us consider the following {\it ansatz} for the solution to the Knizhnik-Zamolodchikov equation
\be
|{\mathcal F}_{j_1,j_2,j_3,j_4}(x,z)|^2=\int d\mu (j){\mathcal G} _{j_{1},j_{2,}j,j_{3},j_{4}} (z;x) \times \bar {{\mathcal G}} _{j_{1},j_{2,}j,j_{3},j_{4}} (\bar z;\bar x)  \label{antonia}
\ee
where the formal sum $\int d \mu (j)$ is given by the measure \cite{teschner} 
\begin{eqnarray}
\int_{{\mathcal C}}dj\frac{C(j_{1},j_{2},j)C(j,j_{3},j_{4})}{B(j)}{\mathcal G}
_{j_{1},j_{2,}j,j_{3},j_{4}}(z;x)  \times \bar {{\mathcal G}} _{j_{1},j_{2,}j,j_{3},j_{4}} (\bar z;\bar x) \label{carita}
\end{eqnarray}
This is the factorization {\it ansatz}: where, as mentioned, $C(j_{1},j_{2},j_{3})$ and $B(j_{1})$ are given by the structure constants and the reflection coefficient respectively, and
where ${\mathcal C}=\frac{1}{2}+i\R$. The integration over the contour ${\mathcal C}$ turns out to be redundant for a monodromy invariant solution since such particular linear combination is invariant under Weyl reflection $\sigma$ and the contour transforms as conjugation $\sigma: {\mathcal C} \rightarrow {\mathcal C}^*=-{\mathcal C}$, see (\ref{below}) below.

Let us expand the chiral\footnote{we use the nomenclature {\it chiral} to refer to the {\it holomorphic part}.} conformal blocks as 
\be
{\mathcal G}
_{j_{1},j_{2,}j,j_{3},j_{4}} (z;x)=z^{h_{j}-h_{j_{1}}-h_{j_{2}}}x^{j-j_{1}-j_{2}}%
\sum_{n=0}^{\infty }{\mathcal G} _{j_{1},j_{2,}j,j_{3},j_{4}}^{(n)}(x)z^{n} \label{niato}
\ee
Here, $j$ acts as an {\it internal} index which labels different solutions to the differential equation; but in the {\it stringy} interpretation it is feasible to assign to this index the physical meaning of parametrizing the intermediate states interchanged in a given four-point process. Then, an integration on $j$ is performed in order to include all the contributions of the conformal blocks (\ref{antonia}) to the four-point function \cite{teschner}.

By replacing (\ref{niato}) into KZ equation we find the following hypergeometric equation for the leading term in the $z$ power expansion
\[
x(x-1)\frac{\partial ^{2}}{\partial x^{2}}{\mathcal G}
_{j_{1},j_{2,}j,j_{3},j_{4}}^{(0)}(x)+(2j-(2j-j_{1}-j_{4}+j_{2}+j_{3}+1)x)%
\frac{\partial }{\partial x}{\mathcal G} _{j_{1},j_{2,}j,j_{3},j_{4}}^{(0)}(x)+
\]
\[
-(j^{2}+j(j_{3}+j_{2}-j_{1}-j_{4})+j_{1}j_{4}+j_{2}j_{3}-j_{1}j_{3}-j_{2}j_{4}){\mathcal G} _{j_{1},j_{2,}j,j_{3},j_{4}}^{(0)}(x)=0
\]
Teschner argued in \cite{teschner} how to obtain the other contributions of the series by a recursive equation ($i.e.$ the terms ${\mathcal G}_{j_{1},j_{2,}j,j_{3},j_{4}}^{(n>0)}(x)$ are uniquely defined up to the choice of a particular solution ${\mathcal G} _{j_{1},j_{2},j,j_{3},j_{4}}^{(0)}(x)$).

\subsection{Monodromy invariants}

Then, in the $2j\notin \Z$ case, the general solution of this leading order can be written as 
\begin{eqnarray}
{\mathcal G} _{j_{1},j_{2,}j,j_{3},j_{4}}^{(0)}(x) &\times& \bar{{\mathcal G}} _{j_{1},j_{2,}j,j_{3},j_{4}}^{(0)}(\bar x) =\left|
F(j-j_{1}+j_{2},j+j_{3}-j_{4},2j,x)\right| ^{2}+ \nonumber  \\
&+&\lambda _{j_1,j_2,j,j_3,j_4}\left|
x^{1-2j}F(1-j-j_{1}+j_{2},1-j+j_{3}-j_{4},2-2j,x)\right| ^{2}  \label{asterisco}
\end{eqnarray}
where $F(\alpha ,\beta ,\gamma ,x)$ is the hypergeometric function.

Monodromy invariance at the point $x=1$ demands that
\be
\lambda _{j_1,j_2,j,j_3,j_4}=-\frac{\Gamma ^{2}(2j)\Gamma (1-j-j_{1}+j_{2})\Gamma
(1-j+j_{3}-j_{4})\Gamma (1-j+j_{1}-j_{2})\Gamma (1-j-j_{3}+j_{4})}{%
\Gamma ^{2}(2-2j)\Gamma (j+j_{1}-j_{2})\Gamma
(j-j_{3}+j_{4})\Gamma (j-j_{1}+j_{2})\Gamma (j+j_{3}-j_{4})}  \nonumber \\
\label{below}
\ee
It is the unique non-trivial monodromy invariant linear combination.

This linear combination vanishes in the point $x=1$. To show this explicitly, let us rewrite $\lambda_{j_1,j_2,j,j_3,j_3}$ as
\begin{eqnarray*}
\left| \frac {\Gamma (\gamma )\Gamma (\gamma -\alpha -\beta)}{\Gamma (\gamma -\alpha )\Gamma (\gamma -\beta )}\right| ^{2}+\lambda _{j_1,j_2,j,j_3,j_3}\left| \frac{\Gamma (1-\gamma )\Gamma (\gamma -\alpha -\beta )}{\Gamma (1-\alpha )\Gamma (1-\beta )}\right| ^{2}(1-\gamma )^{2} =0
\end{eqnarray*}
being $\alpha = j-j_1+j_2$, $\beta = j+j_3-j_4$ and $\gamma =2j$; which is achieved by using standard formulae of $\Gamma$ functions, {\it e.g.} $\frac{\Gamma (1-x)}{\Gamma (n-x)}=(-1)^{n+1}\frac{\Gamma
(1+x-n)}{\Gamma (x)}$. Thus, taking into account that the hypergeometric function satisfies $F(\alpha ,\beta ,\gamma ,1)=\frac{\Gamma (\gamma )\Gamma (\gamma -\alpha -\beta)}{\Gamma (\gamma -\alpha )\Gamma (\gamma -\beta )}$, we finally recognize the condition ${\mathcal G}^{(0)}(x=1) = 0$. Then, this is vanishing for generic values of $j$ in the point $x=1$, and it is also possible to find a vanishing solution for all points $x$ in a particular value of $j$ since (\ref{below}) satisfies the remarkable property 
\begin{equation}
\lim _{j \rightarrow \frac 12} \lambda _{j_1,j_2,j,j_3,j_4}=\lim _{j \rightarrow \frac {k-1}{2}} \lambda _{s(j_1),j_2,s(j),j_3,s(j_4)}=-1 \label{pupo}
\end{equation}
This equation will acquire importance in the next subsection where we analyse the resonant points of the hypergeometric equation.

In \cite{mo3} it was shown the convenience of considering a different expansion for the solution ${\mathcal G} _{j_{1},j_{2,}j,j_{3},j_{4}}$. The proposal was to expand the solution in the vicinity of the point $w=zx^{-1}=0$ in terms of powers of $x$; namely
\be
{\mathcal G} _{j_{1},j_{2,}j,j_{3},j_{4}} (z;x)=x^{\tilde{h}_{j}-\tilde{h}_{j_{1}}-%
\tilde{h}_{j_{2}}}w^{{h_{j}-h_{j_{1}}-h_{j_{2}}}}\sum_{n=0}^{\infty }\tilde{{\mathcal G}}_{j_{1},j_{2,}j,j_{3},j_{4}}^{(n)}(w)x^{n} \label{elotro}
\ee
and analogously for the antiholomorphic part.

This enables one to analyse, for instance, the monodromy properties in the point $w=1$ ({\it i.e.} $x=z$) for leading orders in the $x \rightarrow 0$ limit.
Thus, we will find useful to study the solution in terms of this expansion with the intention to discuss the analytic structure of KZ equation since the point $x=z$ turns out to be of crucial interest within this context.

In \cite{mo3} this contour
integration was deformed as $s: {\mathcal C} \rightarrow \tilde {\mathcal {C}}=\frac{k}{2}-{\mathcal C%
}=\frac{k-1}{2}-i\R$. This particular integration enables one to observe that, once the integral over $j$ is performed, the solution becomes monodromy invariant at $z=x$ \cite{mo3}.

First of all, let us mention that the following relation holds
\[
\lambda _{s(j_1),j_2,s(j),j_3,s(j_4)} \frac{C(j_{1},j_{2},j)C(j,j_{3},j_{4})}{B(j)}=\frac{%
C(j_{1},j_{2},\tilde \sigma (j))C(\tilde \sigma (j),j_{3},j_{4})}{B(\tilde \sigma (j))}\]
The fixed point of $\tilde \sigma$ Weyl transformation is included in the line parametrized by the indicial exponent of the solution ${\mathcal G}^{(0)}$. Indeed, both terms of the monodromy invariant solution are related by the transformation $j\rightarrow \tilde {\sigma }(j) = k-j-1$. 
Besides, notice that the line $j\in \frac{k-1}{2}+i\R$ is precisely the region of $j$-space which remains invariant, up to the direction of the contour integration, under $\tilde \sigma$. This fact suggests that it is convenient to move the contour integration to this line and, considering the Cauchy theorem, rewrite the leading term of the power expansion of KZ solution as follows
\begin{eqnarray*}
|{\mathcal F}_{j_1,j_2,j_3,j_4} (x,z)|^2&=&\frac{1}{2}\int_{\frac{k-1}{2}+i\R%
}dj|x|^{2(\tilde h_{j}-\tilde h_{j_{1}}-\tilde h_{j_{2}})} |zx^{-1}|^{2(h_{j}-h_{j_{1}}-h_{j_{2}})}%
\frac{C(j_{1},j_{2},j)C(j,j_{3},j_{4})}{B(j)} \times \\
&&\times \left| F(j_1+j_2-j,j_3+j_4-j,k-2j,zx^{-1}) \right| ^{2} \left( 1+ {\mathcal O} \left( z^{-1}x \right) \right)+ 2\pi i \sum _{\{x_i\}}Res _{(x=x_i)}
\end{eqnarray*}
where $\{ x_i\}$ refers to the set of poles located in the region $1 <Re (2j)<k-1$; these poles take the form $j-j_1-j_2\in \N$ if the constraint $\sum _{i=1} ^{4} j_i<k$ is assumed (see \cite{mo3} for the details of the construction).

\newpage

\subsection{Logarithmic solutions to KZ equation}

\subsubsection{$j$-dependent solutions}

The equation (\ref{pupo}), which implies that both terms in the solution of hypergeometric equation become equal, up to a sign, in the limit $j \rightarrow \frac 12$, could suggest that, by introducing an appropriate regularization, derivative terms including logarithms could actually arise in the fixed points of Weyl transformation $\sigma $. 



Then, in order to test this idea, we can try a regularization in order to extract a finite solution in the point $j=\frac 12$. This regularization can be absorbed in the definition of the intermediate state $\Phi _j$. Then, let us regularize the monodromy invariant solution as follows
\[
{\mathcal G} _{j_{1},j_{2,}j,j_{3},j_{4}}^{(0)}(x)\times \bar{{\mathcal G}} _{j_{1},j_{2,}j,j_{3},j_{4}}^{(0)}(\bar x) \rightarrow {\mathcal G} _{j_{1},j_{2,}j,j_{3},j_{4}}^{(0)reg}(x) \times \bar {{\mathcal G}} _{j_{1},j_{2,}j,j_{3},j_{4}}^{(0)reg}(\bar x)=\frac{1}{2j-1}{\mathcal G} _{j_{1},j_{2,}j,j_{3},j_{4}}^{(0)}(x) \times \bar{{\mathcal G}} _{j_{1},j_{2,}j,j_{3},j_{4}}^{(0)}(\bar x)
\]
and, then, by taking $j=\frac{1+\varepsilon }{2}$ in order to parameterize the $j\rightarrow \frac{1}{2}$ limit going over the monodromy invariant family of solutions, we finally obtain
\begin{equation}
{\mathcal G} _{j_{1},j_{2},\frac{1}{2},j_{3},j_{4}}^{(0)reg}(x) \times \bar{{\mathcal G}} _{j_{1},j_{2},\frac{1}{2},j_{3},j_{4}}^{(0)reg}(\bar x)=\left| F(\frac{1}{2}-j_{1}+j_{2},\frac{1}{2}+j_{3}-j_{4},1,x)\right| ^{2}\left( g -\log \left| x\right| \right) + \baseRing {\R }e \left( {\mathcal H}(x,\bar{x} )\right)  \label{lacon}
\end{equation}
with
\begin{eqnarray*}
&&{\mathcal H}(x,\bar{x})=F(\frac{1}{2}-j_{1}+j_{2},\frac{1}{2}+j_{3}-j_{4},1,\bar x) \sum_{r=1}^{\infty }x^{r}\frac{\Gamma (\frac{1}{2}-j_{1}+j_{2}+r)\Gamma
(\frac{1}{2}+j_{3}-j_{4}+r)}{ \Gamma ^2 (r+1) \Gamma
(\frac{1}{2}-j_{1}+j_{2})\Gamma (\frac{1}{2}+j_{3}-j_{4})}\times  \\
\times  &&( \psi (\frac{1}{2}-j_{1}+j_{2}+r)+\psi (\frac{1}{2}-j_{4}+j_{3}+r)-\psi
(\frac{1}{2}-j_{1}+j_{2})-\psi (\frac{1}{2}-j_{4}+j_{3})-2\psi (r+1)+2\psi (1)) 
\end{eqnarray*}
where $g $ is the Euler-Mascheroni constant and $\psi $ is the Euler
function defined in terms of gamma functions in the following way
\[
\psi (x)=\frac{d}{dx}\log \Gamma (x)
\]
Now, the question is whether such normalization required to find a non-vanishing result (\ref{lacon}), is feasible in this theory or it is not. The point is that the answer to this question turns out to be no. In fact, that regularization would correspond to a change in the measure $d\mu (j)$ (see below) which, certainly, would lead to spoil certain properties as crossing symmetry of correlation functions. Then, such a freedom does not exist. 

We arrive in this way at the remarkable result that this $j$-dependent logarithmic solution does not contribute to the four-point function in $AdS_3$. 


To be more precise, let us explain in more detail why the solution vanishes at $j=\frac 12$. First, we observe that two linear independent solutions to KZ equation for values of $j$ generic enough are given by
\begin{eqnarray}
\ \ \ F(j-j_1+j_2,j-j_3-j_4,2j,x) \ , \ \ x^{1-2j}F(1-j-j_1+j_2,1-j-j_3-j_4,2-2j,x)  \label{firulete111}
\end{eqnarray}
which implies that the monodromy invariant solution is given by (\ref{asterisco}). Then, taking into account (\ref{pupo}) and the fact that both terms in (\ref{firulete111}) coincide in the limit $2j \rightarrow 1$, we eventually see that the solution identically vanishes at $j=\frac 12$. Of course, it is also due to the fact that measure (\ref{carita}) does not divenge at this precise point.

Another way to argue that this logarithmic solution is not relevant for this process is by simply noticing that the logarithms appear just at a point in the $j$-space and, because the four-point function involves the integral on $j$, this set of null measure is not contributing to the integral at all. {\it i.e.} these solutions are {\it swept away} when the integration over the $j$-space is performed.


\bigskip

Even though we just remarked that the change of the normalization which is required to find a non-vanishing solution at $j=\frac 12$ is not possible because it would imply a non-well behaved measure $d\mu (j)$, let us make a brief comment on the mechanism which would lead to the logarithmic solution (\ref{lacon}) if the {\it wrong} normalization is considered. The point is again that, consequently with the argument presented above, in order to have a non-vanishing contribution at $j \rightarrow \frac 12$ we should have a divergence (single pole) in the measure $d\mu (j)$ at that point; and this divergence would compensate the vanishing result in the numerator leading to find a finite remnant as in (\ref{lacon}). In order to achieve this, we should consider the change
\begin{equation}
d\mu (j) \rightarrow (2j-1)^{-1} d\mu (j)   \label{arroba}
\end{equation}
in (\ref{carita}). Then, since $d\mu (j)$ is composed by the quotient of a pair of structure constants and the reflection amplitude, we see from (\ref{carita}) that (\ref{arroba}) precisely corresponds to a change in the normalization of the vertex $\Phi _j$ representing the intermediate states in the factorization. This agrees with the observation made in \cite{gastonlog}, where is was pointed out that a change in the normalization $\Phi _j \rightarrow \frac {1}{2j-1} \Phi _j$  is required to find a non-vanishing functional form for the vertex operators in the limit $2j \rightarrow 1$. Moreover, we observe from this analysis that the mentioned regularization of $\Phi _j$ and $d\mu (j)$ is consistent with the one required to find the functional form of the prelogarithmic vertex operator $j=\frac 12$ in the realization of the $\hat {sl(2)}_k$ algebra \cite{gastonlog}. To see that, for instance, one can observe that in the free field description of $SL(2,\R )_k$ theory this normalization leads to a vertex operator of the form $\sim \phi e^{-\frac {\phi }{\sqrt {2k-4}}}$ which generates logarithms because of the presence of the additional linear factor $\sim \phi $ in the limit $j \rightarrow \frac 12$. Then, this explain the appearance of logarithmic contributions to the OPE.

\bigskip

We emphazise that the {\it well behaved} normalization (measure) does not enable these (non-vanishing) logarithmic solutions at $j=-\frac 12$. However, within the context of the conformal field theory structure, it could be interesting to study these. Thus, we find convenient to mention other $j$-dependent explicit solutions in the appendix C as examples. In the following, we will also study the $j$-independent conditions for the arising of logarithmic solutions which do encode physical information about the scattering processes.

\bigskip

On the other hand, analogous considerations to those discussed above would hold for solutions of the form $\tilde {\mathcal G}^{(0)} (x)$ in the fixed point of $\tilde \sigma$, which is located on the opposite side of the unitarity segment. {\it i.e.} we could analyse in similar way the logarithmic singularities arising at the resonant points for the {\it ansatz} (\ref{elotro}). In terms of that expansion, the zero mode of the chiral contribution is given by 
\begin{eqnarray}
\tilde{{\mathcal G}}_{j_{1},j_{2,}j,j_{3},j_{4}}^{(0)}(w)={\mathcal G} _{s(j_{1}),j_{2},s(j),j_{3},s(j_{4})}^{(0)} \left( zx^{-1} \right)  \label{entreomegas}
\end{eqnarray}
ans analogous expression holds for its antiholomorphic part.

And similar discussion to that presented above holds in this case: Both terms in $\tilde{{\mathcal G}}_{j_{1},j_{2,}j,j_{3},j_{4}}^{(0)}$ are
mapped one into the other by $\tilde{\sigma}$ Weyl transformation and these become equal, up to a sign, in the 
$j\rightarrow \frac{k-1}{2}$ limit, where $\tilde{{\mathcal G}}%
_{j_{1},j_{2,}j,j_{3},j_{4}}^{(0)}$ vanishes. Then, it is possible to
regularize $\tilde{{\mathcal G}}_{j_{1},j_{2,}j,j_{3},j_{4}}^{(0)}$ as $\tilde{%
{\mathcal G}}_{j_{1},j_{2,}j,j_{3},j_{4}}^{(0)}\rightarrow \tilde{{\mathcal G}}%
_{j_{1},j_{2,}j,j_{3},j_{4}}^{(0)reg}=\frac{1}{2 s (j)-1}\tilde{{\mathcal G}}%
_{j_{1},j_{2,}j,j_{3},j_{4}}^{(0)}$ in order to obtain a finite solution of the
form $\tilde{{\mathcal G}}_{j_{1},j_{2,}j,j_{3},j_{4}}^{(0)reg}(w)={\mathcal G} _{s(j_{1}),j_{2},s(j),j_{3},s(j_{4})}^{(0)reg}(zx^{-1})$ which contains a term of the form $\sim \log \left| z\right| -\log \left|
x\right| $ as solution of KZ equation for $\frac {k-1}{2}$.

\subsubsection{Indicial equation and $j$-independent solutions}

Let us move to the $j$-independent logarithmic solutions, {\it i.e.} $j$-independent sufficient conditions for the appearance of logarithmic terms in the solutions to the KZ equation.

Up to now, we considered two different proposals for the solutions to the KZ equation, and we observed that each of them leads to obtain an hypergeometric equation. Hence, {\it a priori}, we would expect six\footnote{{\it i.e.} three branch points for each hypergeometric equation.} different resonant points appearing in the differential equation. In fact, we identified two of these with fixed points of Weyl symmetry $\sigma $ and $\tilde \sigma$; the remaining will characterize resonance configurations of the external momenta, being independent of the index $j$.

We begin with the solution ${\mathcal G}^{(0)} (x)$. In that case, the hypergeometric equation presents singular points in $x=(0,1,\infty )$, where the generic solutions take the schematic form ${\mathcal G}^{(0)}(x) \sim (x-x_0)^\eta \sum _{n\in \N} a_n (x-x_0 )^n$, where $\eta $ is the indicial exponent. On the other hand, if the roots of the indicial polynomial become double roots, then logarithmic solutions of the following form appear: ${\mathcal G}^{(0)}(x) \sim \eta (x-x_0)^\eta  \sum _{n\in \N} a_n (x-x_0 )^n \log (x-x_0)$. In order to study the appearance of such logarithmic terms in the solutions it is necessary to analyse the indicial polynomial
\be
\eta^2+({\mathcal P}_0-1)\eta +{\mathcal Q}_0=0
\ee 
where ${\mathcal P}_0$ and ${\mathcal Q}_0$ are the complex values which come from evaluating the following rational functions in each singular point $x_0$  
\be
{\mathcal P}(x,x_0 )={(x-x_0)\over x(1-x)} \left( 2j +(j_1-j_2-j_3+j_4-2j)x \right)
\ee
\be
{\mathcal Q}(x,x_0 )={(x-x_0)^2\over x(1-x)} \left( (j_1-j_2-j)(j_3-j_4+j) \right)
\ee
respectively.

Since we are interested in the interpretation of the logarithmic singularities of the four-point function, we address the question about the solutions to the equation
\be
\eta _+(x_0)-\eta _-(x_0)=0 \label{turbo}
\ee
This enables us to study the conditions for which logarithms in $x$ variable appear as solutions to the KZ equation representing correlators of boundary conformal field theory.

Then, for $x_0=0$ we have $2\eta _\pm={1-\gamma \pm |1-\gamma |}$, while for $x_0=1$ we have $2\eta _\pm={\gamma -\alpha -\beta \pm |\gamma -\alpha -\beta |}$ and for the point at infinite $x=\infty $, $2\eta _\pm={\alpha +\beta \pm|\alpha -\beta |}$. Thus, we see that $\eta$ becomes a double root in the cases $j=\frac 12$, $j_1+j_4=j_2+j_3$ and $j_1+j_3=j_2+j_4$ for the points $x_0 =(0,1,\infty )$ respectively. For these conditions, the solution takes the general form
\be
{\mathcal G}_{j_1,j_2,j,j_3,j_4}^{(0)}(x)=W_1(x) + \lambda W_2(x)
\ee
being the functions $W_1$ and $W_2$ given by the following expressions for each case. For $j=\frac 12$ we find
\be
W_1(x)=\sum_{n=0}^\infty a_nx^n, \ \ \ \ W_2(x)= W_1(x)\log (x)+\sum_{n=0}^\infty {\partial a_n(\eta)\over \partial \eta}|_{j=\frac 12}x^n,   \label{d19}
\ee
which is precisely the solution (\ref{lacon}) we studied above.
Analogously, we find a $j$-independent condition; namely, for $j_1+j_4=j_2+j_3$ we have
\be
W_1(x)=\sum_{n=0}^\infty a_n(x-1)^n, \ \ \ \ W_2(x)=W_1(x)\log (x-1)+\sum_{n=0}^\infty {\partial a_n(\eta )\over\partial \eta }|_{j_1+j_4=j_2+j_3}(x-1)^n,
\ee
and for $j_1+j_3=j_2+j_4 $ we obtain similar logarithmic contributions in terms of $x^{-1}$.

When both roots $\eta _{\pm}$ of the indicial polynomial differ by an integer number we find also resonances and the solutions also contain singular points in the corresponding conditions. In some of these cases logarithmic terms could also appear, otherwise the following functional form is obtined for $\eta _+-\eta _-\in \N$
\be
W_2(x)=\sum_{n=0}^\infty c_n(x-x_0)^{\eta _{-}+n}-{\partial W_1(x,\eta)\over\partial \eta}|_{\eta=\eta_{+}} \label{estapibe}
\ee
These poles are, of course, also located at the points $x_0=\{0,1,\infty \}$ and the corresponding conditions are
\be
2j \in { \Z_{\neq 1}} \ , \ \ \ j_1+j_4-j_2-j_3\in \Z_{\neq 0}; \ , \ \ \   j_1+j_3-j_2-j_4\in \Z_{\neq 0}  \label{turboa}
\ee 
Besides, one could be intrigued because of the behavior of the $j$-dependent condition at $x=0$ in (\ref{d19}) which does not lead to recover the condition $j_1+j_2=j_3+j_4$ as one would naively expect. However, we find this condition in the singular point $z^{-1}x \sim 0$, since completely analogous treatement of the case (\ref{elotro}) leads us to obtain the following resonances for the singular points $w=zx^{-1} =\{0,1,\infty \}$
\be
2j \in {k-\Z_{\neq 1}}; \ \ \  k-(j_1+j_2+j_3+j_4)\in \Z_{\neq 0}; \ \ \ j_1+j_2-j_3-j_4\in \Z_{\neq 0}  \label{turboe}
\ee
respectively. Notice that conditions (\ref{turboe}) are not a surprise since these can be derived from (\ref{turboa}) by noting the relation (\ref{entreomegas}). This manifestly shows the relation between the {\it new} condition $k=j_1+j_2+j_3+j_4$ and the existence of an additional singular point in $x=z$; (\ref{entreomegas}) precisely connects this additional singularity with the behavior at $x=1$.

We observe that logarithmic solutions are also found for the condition $j_1+j_2=j_3+j_4$. 

On the other hand, it is possible to verify that the conditions and pole contributions holding for those cases where the roots of the indicial polynomial differ in an integer number $n$ are constrained by unitarity and locality bounds as $|n|<k-2$.

The condition at $j={k-1\over 2}$ is similar to the point $j=\frac 12$ analysed before; then we see that the integral over the contour on $j$-space dilutes the logarithmic singularities which appear located in the fixed points of both Weyl transformation $\sigma$ and $\tilde \sigma$. Let us emphazise that, in addition, we have also found logarithmic terms at $ k=j_1+j_2+j_3+j_4$ in the vicinity of $z=x$.

\subsection{Dynamics in $AdS_3$ space-time}

Now, it would be our intention to interpret the logarithmic solutions of KZ equation presented in the last subsection or, more specifically, to analyse whether the solutions we have found are consistent with the standard interpretation of such type of singularities in the usual studies of the $AdS/CFT$ correspondence.

\subsubsection{The pattern}

First, we obtain from (\ref{turboa}) and (\ref{turboe}) the following pattern: {\it e.g.} we find logarithmic solutions of the form $\sim \log (x-1)$ if $j_1+j_4=j_2+j_3$, while we find singular points of the form $\sim (x-1) ^{j_1+j_4-j_2-j_3}$ if  $j_1+j_4<j_2+j_3$ (and regular otherwise). This pattern was also observed by D'Hoker, Mathur, Matusis and Rastelli in reference \cite{ras4}; where it was remarked that these are precisely the singularities expected from the contributions of composite operators $:{\mathcal O}_2{\mathcal O}_3:$ to the operator product expansion of two fields ${\mathcal O}_1{\mathcal O}_4$.

As it is known, logarithmic contributions can be generated in the CFT by higher order perturbative corrections to the anomalous dimension and by operator mixing. Up to higher corrections, the generic two-particle states of the form $: \partial ^n {\mathcal O}_{j_{\mu }}{\mathcal O}_{j_{\nu }}:$ have conformal dimensions $\Delta =j_{\mu }+j_{\nu }+n$; then, {\it a priori}, it could be feasible to have in the theory different states with the same conformal weight. When such a degeneration occurs, the perturbative corrections can contribute to the leading term of the perturbative expansion. This phenomenon explains the appearance of logarithms $\sim \log (x-1)$ in the correlators, see \cite{ras4,mo3}. 

In \cite{mo3}, Maldacena and Ooguri identified a dependence $\sim x^{j_3+j_4-j_1-j_2}$ in the factorization limit of the four-point function in $AdS_3$ for the region of integration $|z|>1$ in the world-sheet. This dependence was also interpreted as a two-particle contribution.

\subsubsection{On logarithms and $AdS/CFT$}

The appearance of logarithms in different terms of the four-point functions was explained with expertness by the authors mentioned above and many others in references \cite{liu,ras1,ras2,ras3,ras4}.

The authors of \cite{ras1} discussed the issue of logarithmic singularities and their significance for the operator product expansion in the conformal field theory. In that work, the four-point function of chiral primaries corresponding to the contents of scalar fields in type IIB supergravity was analysed in terms of the $AdS_5 / CFT_4$ correspondence. Logarithmic singularities were found in the four-point functions involving chiral states corresponding to the tree-level interaction of dilaton-axion fields. 


Liu showed in \cite{liu} that similar logarithmic singularities appear in the scalar exchange diagrams in $d$-dimensional $AdS _d$ space if certain relations between the conformal dimension of the operators are satisfied. It is mentioned in his article that it would be possible that when adding up all the diagrams contributing to the four-point correlation function the logarithmic divergences cancel\footnote{see also the related work \cite{sanjay}.}. The point is that, even though the chiral primary operators are protected by supersymmetry, mixing effects of two-particle states could appear in the operator product expansion. It was pointed out that the particular pattern of the OPE of descendant operator suggests the mismatch may be due to some mixing effects among different operators. Within the context of this interpretation, it was also remarked that logarithmic divergences appear when the quantum numbers of descendant operators degenerate and, thus, the cancellation of leading terms could lead to higher orders containing derivatives which are responsible for the logarithmic contributions. This is a typical issue in the resonance phenomena.

On the other hand, in reference \cite{ras2} it was shown that logarithmic singularities appear in the four-point functions at subleading orders. The authors first mention the presence of logarithmic terms in the correlation functions involving four states in the coincidence limit of the inserting points; and they signal that it seems to be a characteristic aspect of this class of process in $AdS$ spacetimes. Then, it was emphasized that the logarithmic singularities do not indeed occur in the complete computation of four-point functions. As it is remarked in that paper, referring to an argument attributed to Witten, logarithmic contributions could be due to mixings and perturbative $1/N$ corrections of the exchanged operators. For instance, in the case of the $AdS_5 / CFT_4$ correspondence, renormalization could affect non-chiral composite operators representing two-particle states even though the conformal dimension of the chiral operators are protected.

The logarithmic singularities found in the three dimensional case fit into the same pattern. We can summarize our logarithmic solutions as follows.

\subsubsection{$j$-dependent solutions at fixed points of Weyl transformations}

We have found logarithmic solutions for Knizhnik-Zamolodchikov equation which are located in different points of moduli space. First of all, we obtained logarithmic terms of the form $\sim \log |x|$ which appear for
\begin{eqnarray}
j= \frac 12 \ , \ \ \ \ \ \ j=\frac {k-1}{2}
\end{eqnarray}
and are located in the vicinities of the singular points $x=0$ and $zx^{-1}=0$ respectively. These are $j$-dependent conditions and, as mentioned, do not contribute once the integral over the momentum space is performed. As we explained in section 2.3, at these particular points the solutions of KZ equations written in terms of the factorization {\it ansatz} vanish due to (\ref{pupo}) and the fact that both terms in the solutions (which are independent for generic values) degenerate and cancel mutually. Then, the logarithmic solutions at $j = \frac k4 \pm \frac k4 \mp \frac 12$ vanish because of the precise normalization of the measure $d\mu (j)$ (recall the discussion of section 2.3; see also the logarithmic solutions mentioned in appendix A in order to learn more about a similar mechanism).

\subsubsection{$j$-independent solutions and resonance conditions}

We also find $j$-independent conditions for logarithmic contributions. Some of these are the resonant cases
\begin{eqnarray}
j_1+j_2=j_3+j_4 \ , \ \ \ \ \ j_1+j_3=j_2+j_4 \ , \ \ \ \ \ j_1+j_4=j_2+j_3  \label{poi}
\end{eqnarray}
and are suitable of physical interpretation which describes that composite operators are contributing to the operator product expansion of external states.

These correspond to the expansion around the points $xz^{-1}=0$, $x=\infty $ and $x=1$. We can try an interpretation for the logarithms arising in the conditions like $j_1+j_2=j_3+j_4$ as yielding from higher order corrections which are contributing at this level because of certain cancellation induced for the degeneracy in the conformal dimensions of the operators. This interpretation is supported by the fact that the logarithms precisely appear when the external quantum numbers are tuned in this precise way; in this case there exists degeneration in the two-particle spectrum since the operators $:{\mathcal O}_1{\mathcal O}_2:$
 and $:{\mathcal O}_3{\mathcal O}_4:$ have the same conformal dimension, which, up to higher corrections, is $j_1+j_2$ and coincides with the sum of the dimensions of the operators ${\mathcal O}_1$ and ${\mathcal O}_2$. A related effect was also signaled in \cite{mo3} as the explanation of certain singularities in three-point functions, when it was discussed that mixing between single and multiple-particle states can occur in higher order corrections in the BCFT.

\subsubsection{On logarithmic conformal field theory}

The logarithmic solutions to the $SL(2,\R )_k$ KZ equation were also discussed in the literature in the context of logarithmic conformal field theory (LCFT) and many other interesting topics. In reference \cite{nichols} exact and asymptotic solutions containing logarithms both in the coordinates $z$ and $x$ were found. Also in \cite{log} solutions for interesting examples were explicitly described. For instance, logarithmic contributions of the form $\sim \log z +\frac {k-2}{2j_3-1} \log x$ and $\sim \log (1-z) +\frac {k-2}{2j_3-1} \log (1-x)$ were reported for particular configurations satisfying $j_1+j_2=j_3+j_4$ or permutations when, at least, a pair of these indices vanish ({\it e.g.} $j_1=j_4=0$). The implications of considering such solutions of the KZ equation to the structure of operator product expansion and the Jordan blocks of the Kac-Moody and Virasoro algebras were studied in the mentioned papers and references therein.


In the literature there exists an extended {\it bestiary} of logarithmic solutions to KZ equation. In several works, is was suggested that these are related with some intrinsic property of non-compact WZNW model, proposing this as an example of LCFT. 

However, even though one could accept the feasible consistency of attributing the appearance of logarithmic terms to the LCFT nature of this model, it is also true that this explanation for certain logarithms arising in the four-point functions, by itself, turns out to be unsatisfactory. This is because one would like to have a unified vision of the logarithmic contributions to the correlator structure in the context of $AdS_d/CFT_{d-1}$ correspondence. And then, since similar logarithmic terms were obtained, for instance, in the five dimensional case (were the dual $CFT_4$ involved is the ${\mathcal N}=4$ SYM which is unitary and thus can not represent an example of LCFT), it would be preferable to try another interpretation of some of the logarithms appearing in the four-point functions in $AdS_3$ too.

With this motivation, logarithmic singularities arising in the solutions of the $SL(2,\R )_k$ Knizhnik-Zamolodchikov equation representing four-point functions on $AdS_3$ were discussed in the first part of this paper. And it was observed that the analytic structure found here is suitable to be incorporated into the usual features of the $AdS/CFT$ correspondence previously studied. The logarithmic terms appearing were classified and it was argued that some of these can be related to degeneration in the conformal dimension, presenting a pattern which is similar to the one studied in previous computations within the context of higher dimensional examples. These resonant points, where logarithmic solutions arise, can be interpreted as the spectrum of two-particle contribution to scattering processes.

\subsubsection{$j$-independent solutions and finite $k$ effects}

In addition, singularities corresponding to non-perturbative effects were found. These appear when the following condition is satisfied
\begin{eqnarray}
j_1+j_2+j_3+j_4= k   \label{poi2}
\end{eqnarray}
and are located in the region $x \sim z$.

This is a condition which could not be seen in the framework of the supergravity description of the four-point correlation functions; it corresponds to finite $\alpha '$ regime\footnote{recall the relation $k \sim \frac {l^2_{AdS}}{l^2_{s}}$ holding for the $AdS$ radius $l_{AdS}$ and the string typical lenght $l_{s}$.}.

This is a point that deserves attention; {\it i.e.} we observe the appearance of logarithmic singularities located in $x=z$ for $k=j_1+j_2+j_3+j_4$. Divergences located in this point of the momenta space were signaled in recent references, see for instance \cite{ponsot,mo3,teschner}. In fact, it is feasible to notice, {\it e.g.} from (\ref{estapibe}), that there exist solutions for the four-point functions of the form 
\be
{\mathcal F}_{j_1,j_2,j_3,j_4}(z;x)\sim (x-z)^{k-j_1-j_2-j_3-j_4} + ... \label{i}
\ee
Indeed, there are several ways to see the arising of such a dependence in four-point function. For example, it can be observed directly from our equation (\ref{estapibe}) and can be also found if a general power dependence is assumed as an {\it ansatz} to solve the KZ equation as made in \cite{mo3}. This power expansion can be obtained directly by noting the following expansion of the hypergeometric functions involved in the solution 
\be
F(j_{1}+j_{2}-j,j_{3}+j_{4}-j,k-2j,w)=  \frac {\Gamma
(k-2j)\Gamma (k-j_{1}-j_{2}-j_{3}-j_{4})}{\Gamma (k-j-j_{1}-j_{2})\Gamma
(k-j-j_{3}-j_{4})} + ... +
\ee
\be
+ \frac{\Gamma (k-2j)\Gamma (j_{1}+j_{2}+j_{3}+j_{4}-k)}{\Gamma
(j_{1}+j_{2}-j)\Gamma (j_{3}+j_{4}-j)}(x-z)^{k-j_{1}-j_{2}-j_{3}-j_{4}} x^{j_{1}+j_{2}+j_{3}+j_{4}-k} \times \left( 1+{\mathcal O}\left( 1-zx^{-1} \right) \right) \nonumber  \\ 
\label{tesis}
\ee
Analogously, we obtain the dependence $\sim (x-1)^{j_1-j_2-j_3+j_4}$ mentioned before in the coincidence limit $x \rightarrow 1$ by similar analysis. 

Maldacena and Ooguri presented in \cite{mo3} an interpretation of the poles of the functional form (\ref{i}) in terms of instantonic contributions in the world-sheet theory. The dependence (\ref{i}) has been also mentioned in \cite{nichols} when it was pointed out that the four-point function ${\mathcal F}_{j_1,j_2,j_3,j_4}(z;x)$ satisfies the same KZ equation as another solution which is proportional to $(x-z)^{k-j_1-j_2-j_3-j_4} $ $ {\mathcal F}_{s(j_1),s(j_2),s(j_3),s(j_4)}(z;x)$; similar correspondence was studied by Ponsot in reference \cite{ponsot}. In \cite{petko} the general form of the four-point correlators was constructed in terms of powers of $(z-x)$, and a prescription for a Felder-type integration was tried in order to deal with the singularity in $z=x$, see \cite{petko,petko2}.

Another interesting issue, which is related with (\ref{entreomegas}), is the fact that condition $ k=j_1+j_2+j_3+j_4$ can also be written as $ s(j_1)+s(j_2)=j_3+j_4$. Taking into account that the spectral flow automorphism in the sectors $\omega \in \{-1,0,1\}$ is closed among the {\it standard} representations of $SL(2,\R)$ and that there exist states with identical Casimir related by $j \leftrightarrow s(j)$, this could be suggesting that, in some sense, this $k$-dependent condition can be also considered as a resonance condition on equal footing with the other $j$-independent relations. Up to this point, this has to be considered simply as an heuristic argument, but the connection between the different $j$-independent resonant conditions (\ref{poi}) and (\ref{poi2}) acquires substance when the symmetries of the model ({\it v.g.} the spectral flow) are analysed in terms of the Liouville description of correlators in the following section and appendix B.


A first simple observation that could be made is that, as in the case of $s$-transformation, $r$-transformations also leave invariant the set of four $j$-independent resonant conditions (\ref{poi}) and (\ref{poi2}). These are particular cases of a set of more general symmetries discussed in the following.

\section{Knizhnik-Zamolodchikov equation and Liouville theory}

Here, in order to study several important aspects of the solutions of KZ equation, let us make use of a quoted result due to Fateev and Zamolodchikov \cite{fz4p}. In that article, the authors have shown that certain five-point functions of minimal models, which contain a degenerate field, satisfy the KZ equation for the $SU(2) _k$ WZNW model. 

The application of a similar result for WZNW model formulated on $SL(2,\R)$ has been presented in the literature. We use it here to explore the singular structure of solutions to the KZ equation and, principally, to clarify the relation existing between different $j$-independent conditions for the appearance of logarithmic terms. The main goals are basically two: first, we find how to recover the logarithmic solutions found in the previous section from the operator product expansion in the Liouville description. Secondly, we show how the reflection properties of Liouville theory allow to systematically detect symmetries of KZ equation. We do this in appendix B.

\subsection{WZNW correlators from Liouville theory}

The straightforward generalization of the result \cite{fz4p} leads to notice that the BPZ equation satisfied by five-point functions in Liouville CFT coincides with the KZ equation satisfied by four-point functions in the WZNW model on $SL(2,\R )$. This result is a powerful tool which enables to study non trivial features of the four-point functions in the non-rational conformal field theories. For instance, Teschner proved in \cite{teschner} the cross symmetry of the model on the $SL(2,\C)/SU(2)$ homogeneous space from the one holding in Liouville theory; and Ponsot showed in \cite{ponsot} how to write representations for the monodromy of the conformal blocks in the $SL(2,\C)/SU(2)$ by means of the corresponding quantities of Liouville CFT.

\subsubsection{The precise correspondence for the $SU(2)$ case}

The relation pointed out by Fateev and Zamolodchikov asserts that the BPZ equation satisfied by five-point correlators including degenerate fields $\psi _{2,1}$ (or $\psi _{1,2}$) in the minimal models exactly coincides with the Knizhnik-Zamolodchikov equation for four-point functions in the $SU(2)_k$ WZNW model when the following relations between the conformal dimensions $\Delta $ of degenerate operators and the Kac-Moody level $k$ hold
\begin{eqnarray*}
4\Delta_{(2,1)}+2=3(k+2) \ , \ \ \ 4\Delta_{(1,2)}+2=3(k+2)^{-1}
\end{eqnarray*}
for the case $\psi _{2,1}$ (resp. $\psi _{1,2}$). The non-compact case $SL(2,\R )_k$ which is of our interest here is related to the $SU(2)_k$ case by replacing $k \rightarrow -k$, while the degenerate states of minimal models $\psi _{2,1}$ (and $\psi _{1,2}$) are in correspondence with the state(s) $V_{\alpha = -\frac {1}{2b}}$ (and resp. $V_{\alpha = -\frac {b}{2}}$) of Liouville CFT respectively, being $b^{-2}=k-2$. This is consistent with the relation existing between $\Delta_{(2,1)}$ and $\Delta_{(1,2)}$, which implies that the interchange $b \rightarrow b^{-1}$ corresponds to the inversion $k-2 \rightarrow \frac {1}{k-2}$, with fixed point at $b=1$ and $k=3$ (and also $k=1$).

\subsubsection{The setup for the non-compact case}

As we will describe in the following subsections, the non-compact generalization of the correspondence pointed out in \cite{fz4p} is an important result which allows us to write down an explicit form for four-point functions in WZNW model on $SL(2,\R)$ in terms of the five-point functions of Liouville field theory. These five-point correlators include states with null descendants represented by Liouville vertex operators of the form $e^{\sqrt {k-2} \varphi (x)}$ and can be realized in the Coulomb-gas like prescription by inserting screening operators of the form $e^{2(k-2)^{\mp 1/2}\varphi (w)}$.

First, let us consider the following setup for the solution: $h_{34} =-h_{2}-h_{1}+h_{3}-h_{4}$, $h_{14} =-2h_{1}$, $h_{24} =h_{2}-h_{1}-h_{3}-h_{4}$, $h_{23} =h_{4}+h_{1}-h_{2}-h_{3}$ and $J_{34} =j_{1}+j_{2}-j_{3}+j_{4}$, $J_{14} =-2j_{1}$, $J_{24} =j_{1}-j_{2}+j_{3}+j_{4}$, $J_{23} =j_{1}+j_{2}+j_{3}-j_{4}$ and the simple change $j_1 \leftrightarrow j_2$.

With this convention, we can write down the four-point functions ${\mathcal 
A}^{WZNW}_{j_1,j_2,j_3,j_4}$ in terms of the five-point functions ${\mathcal A}^{Liouville}_{\alpha _1,\alpha _2,-\frac{1}{2b},\alpha_ 
3,\alpha _4}$ (including a particular fifth state $\alpha _5 
=-\frac{1}{2b}$) as follows
\begin{eqnarray}
{\mathcal A}^{WZNW}_{j_1,j_2,j_3,j_4} &=&|x|^{-2\alpha _{2}/b}|1-x|^{-2\alpha
_{3}/b}|x-z|^{-2\alpha _{1}/b}|z|^{-4(b^{2}j_{1}j_{2}-\alpha _{1}\alpha
_{2})}|1-z|^{-4(b^{2}j_{3}j_{1}-\alpha _{3}\alpha _{1})}\times \nonumber \\
&&\times f(j_1,j_2,j_3,j_4){\mathcal A}^{Liouville}_{\alpha _1,\alpha 
_2,-\frac{1}{2b},\alpha_ 3,\alpha _4} \label{parcialito}
\end{eqnarray}
where\footnote{%
Notice that, with respect to our previous notation, there exists a difference in nomenclature due to the interchanges $j_{1}\leftrightarrow j_{2}$,  $z_{1}\leftrightarrow z_{2}$ and $\alpha _{1}\leftrightarrow \alpha _{2}$ which we introduce here in order to simplify the comparison with the seminal work \cite{fz4p}.}
\begin{eqnarray*}
2\alpha _{1} =b\left( j_{1}+j_{2}+j_{3}+j_{4}-1\right) \ , \ \ 2\alpha _{i} =b\left( j_{1}-j_{2}-j_{3}-j_{4}+2j_{i}+b^{-2}+1\right)
\end{eqnarray*}
where $i \in \{ 2,3,4 \} $ and $b^{-2}=k-2$. In the expression above $f(j_1,j_2,j_3,j_4)$ represents an overall factor which is (almost) determined by the normalization of structure constants of the theory; we specify it in appendix A.

Then, in this way, we can study the solutions to KZ equation indirectly, by studying correlators on the {\it Liouville side} of the correspondence between both CFT's. With this purpose, let us remind basic aspects of Liouville conformal field theory\footnote{Recently, the interest on Liouville theory has been renewed in the context of two-dimensional string theory \cite{seibergs}, see also \cite{zamo2}.}.

\subsection{Liouville field theory}

Consider the quantum Liouville action
\[
S_L =\int d^{2}z\left( \left( \partial \varphi \right) ^{2} + 4\pi QR\varphi +\mu
e^{2b\varphi }\right)
\]
being
\[
Q=(b+b^{-1})
\]
and where $R$ is the two-dimensional scalar curvature.

The Liouville $\varphi $ field propagator is given by
\[
\left\langle \varphi (z,\bar z)\varphi (w,\bar w)\right\rangle =-\frac{1}{2}\log |z-w|^2
\]
and the stress-tensor and central charge take the form
\begin{eqnarray*}
T &=&-\partial \varphi \partial \varphi +Q\partial ^{2}\varphi \\
c &=&1+6Q^{2}
\end{eqnarray*}
respectively. The primary fields of the following form are important objects in Liouville CFT
\[
V_{\alpha }(z)=:e^{2\alpha \varphi (z)}:
\]
which have conformal dimension
\[
\Delta _{\alpha }=\alpha (Q-\alpha )
\]
Notice that the formula for conformal dimension is invariant under the reflection $\alpha \rightarrow Q-\alpha$; and corresponding states would be related by such conjugation as described below. This is the reflection symmetry which turns out to be of crucial importance in Liouville CFT and, in particular, within the context of our discussion.

Let us consider the composed vertex operator including both contributions $V_{\alpha }$ and $V_{Q- \alpha }$, namely
\begin{equation}
{\mathcal V}_{\alpha }(z)= V_{\alpha }(z) + R_{\alpha} V_{Q- \alpha }(z)  \label{tyt}
\end{equation}
where $R_{\alpha }$ is the Liouville reflection coefficient ({\it i.e.} given by Liouville two-point function) which satisfies \cite{teschnerliouville}
\begin{equation}
R_{\alpha }R_{Q-\alpha }=1 \ , \ \ \ R_{\alpha } {\mathcal V}_{Q-\alpha } (z)={\mathcal V}_{\alpha } (z)  \label{fuba}
\end{equation}
This is related to the asymptotic limit in the free region $\varphi \rightarrow -\infty $ of the Liouville wave function and the fact that both contributions $\alpha $ and $Q-\alpha $ appear in that region.

For the cases $\alpha =\frac Q2 + \epsilon$, the reflection coefficient takes the form 
\[
R_{\frac Q2 + \epsilon }=- \left( \pi \mu  \frac{\Gamma (b^2)}{\Gamma (1-b^2)} \right) ^{-\frac {2\epsilon }{b}} \frac {\Gamma (1+2b\epsilon ) \Gamma (1+2b^{-1}\epsilon )}{\Gamma (1-2b\epsilon ) \Gamma (1-2b^{-1}\epsilon )} 
\]
Hence, we find that the following remarkable property holds
\begin{equation}
\lim _{\alpha \rightarrow \frac Q2} \frac {1}{4 \epsilon} {\mathcal V}_{\frac Q2 + \epsilon} (z) = e^{Q\varphi (z)} \left( \varphi (z)+ \frac {1}{2b} \log \left( \pi \mu \frac {\Gamma (b^2)}{\Gamma (1-b^2)} \right) \right)  \label{hj}
\end{equation}
which defines a primary ({\it puncture}) operator, even though a linear term in $\varphi $ is present. Let us emphasize that (\ref{hj}) has to be understood as the definition of ${\mathcal V} _{\frac Q2}$, namely
\begin{equation}
{\mathcal V} _{\frac Q2} \equiv \lim _{\alpha \rightarrow \frac Q2} \frac {1}{4 \epsilon} {\mathcal V}_{\frac Q2 + \epsilon} (z)   \label{hj2x}
\end{equation} 
because it requires a regularization since the linear term in $\epsilon$ vanishes, see \cite{teschnerliouville}. Then, notice also that the cosmological term of the $c=1$ matter model is a particular case of the above vertex; this is also analogous to the $j= \frac12$ representation in $\hat {sl(2)}_k$ algebra \cite{gastonlog}. We observe from this fact that the reflection coefficient $R_{\alpha }$ presents a resonance in $\alpha = \frac Q2$ and, thus, the corresponding relation (\ref{fuba}) between ${\mathcal V}_{\alpha }$ and ${\mathcal V}_{Q- \alpha }$ breaks down in this particular point; in this case a primary operator with conformal dimension $\Delta _{\frac Q2} = \frac {Q^2}{4}$ is given by $\varphi e ^{Q\varphi }$.
   
In Liouville CFT, there are two different screening operators, namely
\[
S_{\pm }=\int d^2zV_{b^{\pm 1}}(z)=\mu _{\pm} \int d^2 z e^{2b^{\pm 1}\varphi (z)}
\]
being 
\[
\mu _- = \frac {\Gamma (1-b^{-2})}{\pi \Gamma (b^{-2})} \left( \pi \frac {\Gamma (b^2)}{\Gamma (1-b^2)}\mu _+ \right) ^{b^{-2}},
\] 
see \cite{teschnerliouville,zamo2,nos3}. One of these, $S_+$, corresponds to the Liouville cosmological term.

On the other hand, the correlation functions can be written as
\[
{\mathcal A}^{Liouville}_{\alpha _{1},...\alpha _{N}} = \left\langle V_{\alpha 
_{1}}(z_1)V_{\alpha _{2}}(z_2)...V_{\alpha _{N}}(z_N)\right\rangle = \\
\]
\[
= \frac {\mu _+ ^{n_+ + n_- } R_b ^{n_-}}{n_+!n_-!} \left\langle \prod _{\mu =1} ^{N} e^{2\alpha _{\mu }\varphi
(z_{\mu })} \prod^{n_{-}}_{r=1} \int 
d^2v_{r} e^{2b^{-1} \varphi (v_r)}\prod^{n_{+}}_{r=1} \int d^2w_{r} e^{2b 
\varphi (w_r)}\right\rangle
\]
where $n_{\pm }$ refers to the amount of screening operators of the type $S_{\pm }$ required to satisfy the charge symmetry condition yielding from the integration over the zero-mode of $\varphi $ field, namely $\sum _{\mu =1} ^{N} \alpha _{\mu }+n_{+}b+n_{-}b^{-1}=Q$.

Then, the Coulomb gas like realization for the five-point correlators (\ref{parcialito}) leads us to the following necessary condition
\begin{equation}
\sum_{\mu =1}^{4}\alpha _{\mu }-\frac{1}{2b} +n_{+}b+n_{-}b^{-1}=Q  \label{xc43}
\end{equation}
where the contribution $-\frac{1}{2b}$ comes from the presence of the fifth operator representing the degenerate state with $\alpha _{5}=-\frac{1}{2b}$.

Taking into account the relation existing between quantum numbers of both models $\alpha_{\mu }$ and $j_{\mu }$, we finally find
\begin{eqnarray}
n_{+}+n_{-}(k-2)=-2j_{1} \label{usapa}
\end{eqnarray}
and, as a particular case \cite{fz4p}, we have $n_{+}=-2j_{1}$, $n_{-}=0$.

The complete functional form of correlators yields from the insertion of ${\mathcal V}_{\alpha _{\mu}}$ operators.

Further technical steps in the analysis are to establish the integral representation which enables to explicitly realize (\ref{parcialito}), to specify the normalization $f(j_1,j_2,j_3.j_4)$ by basic requirements and to perform checks of the construction by studying monodromy invariance of particular (integrable) cases and the factorization properties. We work out these aspects in detail in the appendix A.

\subsection{Liouville theory description of logarithmic solutions}

Now, we can make use of this dictionary between quantum numbers of both conformal models in order to study the particular configurations leading to logarithmic solutions discussed in section 2. Let us study logarithms as coming from the OPE of degenerate operators and puncture operators.


The main observation is the fact that for each of the four logarithmic conditions (\ref{poi}) and (\ref{poi2}), we obtain that one of the corresponding Liouville states in the five-point correlators becomes represented by a puncture operator $\sim {\mathcal V}_{\frac Q2}$. {\it i.e.} the logarithmic solutions for $\hat {sl(2)} _k$ KZ described in section 2 arise for
\[
j_1+j_2=j_3+j_4 \ , \ \  j_1+j_3=j_2+j_4 \ , \ \  j_1+j_4=j_2+j_3 \ , \ \ j_1+j_2=s(j_3)+s(j_4)
\]
for $x=0$, $x=1$ $x=\infty$ and $x=z$ respectively\footnote{Notice that it is necessary to consider the change in the convention $j_1 \leftrightarrow j_2$ mentioned before for a precise identification. It is consistent with our setup.}; and these are precisely the conditions which correspond to the following particular configurations
\begin{equation}
\alpha _2 =\frac Q2 \ , \ \ \ \alpha _3 =\frac Q2 \ , \ \ \ \alpha _4 =\frac Q2 \ , \ \ \ \alpha _1 =\frac Q2  \label{ert67}
\end{equation}
respectively; where, as mentioned, the corresponding vertex operator ${\mathcal V} _{\alpha _{\mu }}$ developes a linear term in $\varphi $, like in (\ref{hj}). These operators are primary fields\footnote{For a very interesting discussion related with these operators, see the recent work \cite{ultimo}.} in the world-sheet theory as it can be directly observed from the fact that 
\begin{equation}
{\mathcal V}_{\frac Q2} = \frac 14 \frac {\partial }{\partial \alpha} {\mathcal V} _{\alpha \ \ |_{ \alpha =\frac Q2}}   \label{dale7000}
\end{equation}
which can be directly verified by differenciating (\ref{tyt}) respect to $\alpha $.

Then the operator product expansion would take the form\footnote{Recalling that here we have $z_1=z$, $z_2=0$, $z_3=1$ and $z_4=\infty $.} 
\begin{eqnarray*}
{\mathcal V}_{\alpha_{\mu} = \frac Q2} (z_{\mu }) {\mathcal V}_{\alpha _{\nu }} (z_{\nu }) \sim \sum _{\{ \rho \} }  \frac {\partial}{\partial \alpha _{\mu}}  C_{\rho \mu \nu } |z_{\mu }-z_{\nu }|^{2\Delta _{\alpha _{\rho }}-2\Delta _{\alpha _{\mu }}-2\Delta _{\alpha _{\nu }} } \{ {\mathcal V}_{\alpha_{\rho } } \}  \ _{|\alpha _{\mu }=\frac Q2} \ +...
\end{eqnarray*}
which generates a term proportional to $ (Q -2 \alpha _{\mu }) \log |z_{\mu }-z_{\nu }|^2$ because of the presence of $\sim \frac {\partial}{\partial \alpha _{\mu}} \Delta _{\alpha _ {\mu}}$. Thus, such logarithmic contributions containing two $z's$ to the OPE vanish when evaluated at $\alpha _ {\mu}=\frac Q2$. However, rather different contributions stand in the OPE with the degenerate field ${\mathcal V}_{-\frac {1}{2b}} (x)$ since in such a case the exponent in the short distance behavior is linear in $\alpha _{\mu}$, namely 
\begin{eqnarray}
{\mathcal V}_{\alpha_{\mu} = \frac Q2}  (z_{\mu })  {\mathcal V}_{-\frac {1}{2b}} (x) \sim \frac {\partial}{\partial \alpha _{\mu}}  C_{-}  |z_{\mu }-x|^{1+b^{-2}- b^{-2}(1+b^2- 2\alpha _{\mu }b)}  {\mathcal V}_{\alpha_{\mu} - \frac {1}{2b}}  + \nonumber \\
 + \frac {\partial}{\partial \alpha _{\mu}}  C_{+}  |z_{\mu }-x|^{1+b^{-2}+ b^{-2}(1+b^2- 2\alpha _{\mu }b)}  {\mathcal V}_{\alpha_{\mu} + \frac {1}{2b}} \label{p900}
\end{eqnarray} 
where \cite{teschnerliouville, teschnerotra}
\begin{equation}
\frac {C_+ (\alpha _{\mu })}{C_- (\alpha _{\mu })} = -\pi \mu _- \frac {\gamma (b^{-1}(2\alpha _{\mu } -b^{-1})-1) }{\gamma (-b^{-2}) \gamma (2\alpha _{\mu }b^{-1})}  \label{unak}
\end{equation}
being $\gamma (x) = \Gamma(x) / \Gamma (1-x)$ and $C_-=1$. 

This generates the contributions 
\begin{eqnarray}  
\sim &&  C_{-} \ 2b^{-1} |z_{\mu }-x|^{1+b^{-2}- b^{-2}(1+b^2- 2\alpha _{\mu }b)} \log |z_{\mu } - x| {\mathcal V}_{\alpha _{\mu }-\frac {1}{2b}} \ - \nonumber \\
 - && C_{+} \ 2b^{-1} |z_{\mu }-x|^{1+b^{-2}+ b^{-2}(1+b^2- 2\alpha _{\mu }b)} \log |z_{\mu } - x| {\mathcal V}_{\alpha _{\mu }+\frac {1}{2b}} + ... \ \label{p06}
\end{eqnarray} 
Thus, taking into account the factor $\prod ^{4}_{\mu =1} |z_{\mu }-x|^{-2\alpha _{\mu }b}$ standing in (\ref{parcialito}), we observe that the two exponents in (\ref{p06}) 
\begin{equation}
1+b^{-2}\pm b^{-2}(1+b^2- 2\alpha _{\mu }b)-2\alpha _{\mu}b  \label{g67}
\end{equation}
vanish for the value $\alpha _{\mu }=\frac Q2$ cancelling the power dependence and leading to the exact $x$-dependent logarithmic contribution found in section 2. This is, indeed, a check of the logarithmic solutions inferred by solving the factorization {\it ansatz}.

Besides, at this point, one should be wonder about whether a cancellation of both logarithmic terms in (\ref{p06}) occurs or not.

In order to answer to this question in a concise way, we have to consider two different aspects: First, we need to verify that the coefficients $C_+$ and $C_-$ do not cancel mutually at the particular point $\alpha _{\mu } = \frac Q2$ since, due to (\ref{fuba}), we observe that
\[
{\mathcal V}_{\frac Q2 - \frac {1}{2b}} (z) =  R_{\frac Q2 - \frac {1}{2b}} {\mathcal V}_{\frac Q2 + \frac {1}{2b}} (z) 
\]
and, thus, both fields appearing in (\ref{p900}) are proportional to each other and such a cancellation could actually occur. Furthermore, we find that this risk of cancellation actually exists due to the fact that the limit
\[
\lim _{\alpha _{\mu } \rightarrow Q/2} \frac {C_+ (\alpha _{\mu })}{C_- (\alpha _{\mu })} = \lim _{\alpha _{\mu }\rightarrow Q/2} \frac {\pi \mu _- b}{Q-2\alpha _{\mu } }
\]
precisely coincides with the asymptotic behaviour of the reflection coefficient
\[
\lim _{\alpha _{\mu } \rightarrow Q/2} R_{\alpha _{\mu }-\frac {1}{2b}} = \lim _{\alpha _{\mu }\rightarrow Q/2} \frac {\pi \mu _- b}{Q-2\alpha _{\mu } }
\]
However, by using basic functional properties of the reflection coefficient $R_{\alpha }$ ({\it e.g.} the fact that this satisfies $R_{Q/2} = -1$) 
and the presence of the minus sign in the second line of (\ref{p06}) (which comes from the derivative of the exponent (\ref{g67})), we finally find the following non-vanishing remnent contribution in the limit $\alpha _{\mu }\rightarrow Q/2$
\begin{eqnarray*}  
\sim -4 b^{-1} \log |z_{\mu } - x| {\mathcal V}_{\alpha _{\mu }+\frac {1}{2b}} 
\end{eqnarray*} 

Then, it is here where the second aspect enters in the game. In fact, because of the functional form of the normalization factor $f(j_1,j_2,j_3,j_4;k)$ appearing in (\ref{parcialito}) (see (\ref{normaton}) below), an additional singularity also emerges in the limit $\alpha _{\mu} \rightarrow Q/2$. This is due to the fact that the $\Upsilon ^{-1}(2\alpha _{\mu})$ function involved in expression (\ref{normaton}) developes a single pole at the point $2\alpha _{\mu} =b+b^{-1}$ which leads to the form
\begin{equation}
 \lim _{\alpha \rightarrow Q/2} \frac {{\mathcal V} _{\alpha _{\mu }}}{2\alpha _{\mu }-Q} = \frac 14 \lim _{\alpha _{\mu }\rightarrow Q/2} \ \partial _{\alpha _{\mu }} {\mathcal V} _{\alpha _{\mu }}
\end{equation}
arising in the limit of the above equation (instead simply ${\mathcal V} _{Q/2}$). This resembles the connection between (\ref{tyt}) and (\ref{dale7000}), where the factor $(2\alpha _{\mu }-Q)^{-1}$ exaclty corresponds to the $\epsilon ^{-1}$ in (\ref{tyt}). It is important to notice that this is precisely consequent with the definition of ${\mathcal V}_{\frac Q2}$ given in (\ref{hj2x}). Hence, the presence of the normalization factor $f(j_1,j_2,j_3,j_4;k)$ is the responsible of the non-vanishing logarithmic contributions in WZNW correlators and a crucial point for finding the functional from $\partial _{\alpha } {\mathcal V}_{\alpha }$ in the limit $\alpha \rightarrow Q/2$. An important remark within this context is that, even though the apparent cancellation of the logarithmic contribution in the operator product expansion (\ref{p06}) in Liouville theory we discussed above, the presence of the factor (\ref{normaton}) as the one connecting correlators in both conformal models produces that non-vanishing logarithmic contributions appear in the $SL(2,\C )/SU(2)$ WZNW CFT.



Thus, we eventually find non-vanishing logarithmic contributions in the correlator of the $SL(2,\C )/SU(2)$ WZNW model when conditions (\ref{ert67}) are satisfied. Then, we reobtain the logarithmic solutions in this framework (see also\footnote{the correlation functions in Liouville theory have been extensively studied in references \cite{teschnernuevos}} \cite{si}).


Before concluding, let us mention that, even though we made use of this Liouville picture to describe the logarithms arising in the solutions of KZ equation, certain questions could appear due to this. It is the case of the general question about the nature of the logarithmic contributions in the OPE of Liouville correlators which, indeed, is an interesting topic by itself in a more general context. In fact, logarithms in Liouville theory were previously studied in the literature ({\it e.g.} see \cite{si}). And, thus, further study on this puzzle of logarithms in this well-behaved CFT is desirable in order to completely understand the structure of non-rational conformal models. In our case, we observe that there actually exists a relation between the consideration of the Seiberg bound $\alpha _{\mu} < \frac Q2$ and the exclusion of logarithmic solutions. 

In particular, notice that in our formula for four-point function the constraint $\alpha _1 < \frac Q2$ translates into the bound $j_1+j_2+j_3+j_4 < k$ on WZNW four-point function. This shows a connection between the Seiberg bound in Liouville theory and the constraints on four-point correlators in WZNW models discussed in \cite{mo3} within the context of the locality properties of the dual BCFT. It would be interesting to investigate the physical interpretation of such correspondence.

\section{Discussion and Remarks}

We studied the solutions to the Knizhnik-Zamolodchikov equation at the level of four-point function satisfying the factorization {\it ansatz} and containing logarithmic terms in the $(x,\bar x)$ isospin variables. As mentioned, these solutions (corresponding to (\ref{poi}) and (\ref{poi2})) are suitable for physical interpretation in terms of the $AdS/CFT$ correspondence. Indeed, we showed how our solutions reproduce the pattern found in other studies in the topic based on the field theory approximation. The logarithms signal the presence of higher $1/N$ corrections appearing because of composite (two-particle) states in the intermediate channels of the interaction process. This explains the tuning of the external momenta when the logarithmic terms arise in the solution. In addition, solutions related to finite $k$ effects were found.

Then, in section 3, we have seen that the logarithmic solutions are recovered in the Liouville description of the WZNW correlators by noticing the appearance of {\it puncture} operators in correlators and because of the presence of degenerate fields which have a very particular OPE structure. We discussed how the non-vanishing logarithmic contributions in $SL(2,\R )$ WZNW model arise despite certain tuning between the reflection coefficient $R_{\alpha }$ and the constants $C_{\pm }(\alpha )$ in the limit $\alpha _{\mu } \rightarrow Q/2$ which, naively, would lead to conclude the cancellations in the operator product expansion in Liouville five-point function. The normalization connecting correlators in both models and the explicit form of the reflection coeficient are crucial points in this analysis for stablishing the existence of non-vanishing WZNW logarithmic correlators.

First of all, this is a non trivial check of the logarithmic solutions inferred by solving the factorization {\it ansatz} in section 2. On the other hand, we obtained in this way a {\it dual} Liouville picture of the corresponding four-point scattering process in $AdS_3$ and we showed that the resonance conditions (\ref{poi}) and (\ref{poi2}), which are necessary for the appearance of logarithmic singularities in the solutions to the $\hat {sl(2)} _k$ KZ equation, can be viewed as the resonant point of the reflection coefficient in the Liouville operators in the five-point function. 

This identification between observables of both conformal field theories provided us a way to clearly identify the source of the logarithmic contributions in terms of the operator product expansion of Liouville vertices. It also allowed us to find a clear explanation for certain {\it strange} hidden $\Z _2$ symmetries of the KZ equation which translate into simple Liouville reflections on the other side (see appendix B for details). We showed how the reflection symmetry of Liouville theory induces particular automorphisms of the spectral flow symmetry of $\hat {sl(2)} _k$ algebra combined with $\Z _2$ symmetry transformations of the correlators. For instance, it is possible to see in appendix B (see (\ref{n4}) and (\ref{n5})) how the conditions $k=j_1+j_2+j_3+j_4$ and $j_1+j_2-j_3-j_4=0$ are simply related by interchanges ({\it crossing}) of quantum states in Liouville correlators; this turns out to be a useful framework to study conditions of the non-perturbative regime of string theory on $AdS_3$ since the $k$-dependent condition mentioned above is closely related with the singularity located at $x=z$. 


As a concluding remark, we would like to finish this paper by mentioning a point that, in our opinion, is still an open question. In fact, even though the use of the Fateev-Zamolodchikov\footnote{generalized to the non-compact case by Andreev and Teschner.} dictionary between observables of WZNW model and Liouville theory provided us a useful tool to identify and study several aspects, symmetries and constraints of both CFT's, it is also true that a more intuitive interpretation of such correspondence would be desirable in order to completely understand the physical picture. For instance, a concise example is the requirement of a physical interpretation, if it does exist, for the correspondence we have observed between the Maldacena-Ooguri constraint (required for the locality of the BCFT dual of strings in $AdS_3$) and the Seiberg bound on the Liouville {\it side}. Besides, another point that deserves further study is the necessity of a deeper understanding of the physical meaning of the singularity at $z=x$ which, despite the attempts of many experts, remains yet unclear. In particular, the task to find a connection between the logarithmic dependences we have found and the worldsheet instantons pointed out by Maldacena and Ooguri in \cite{mo3} is an interesting topic for investigation in the future.


\[
\]

G.G. is very grateful to J.M. Maldacena and J. Teschner for very useful discussions and illuminating comments. G.G. also thanks M. Farinati and thanks G. Bertoldi, S. Cherkis, Yu Nakayama, L. Rastelli and J. Teschner for suggestions and reading the preliminary manuscript. We thank C. N\'u\~nez for conversations on related subjects. G.G. is supported by Institute for Advanced Study IAS and Fundaci\'{o}n Antorchas. C.S. is supported by Consejo Nacional de Investigaciones Cient\'{\i}ficas y T\'ecnicas CONICET and Universidad de Buenos Aires UBA.

\newpage

\section{Appendix A: Fateev-Zamolodchikov integral representation}

In this appendix we study the integral representation of the five-point function in Liouville CFT and its relation with the four-point function in WZNW model. We fix the normalization and discuss certain technical aspects.

Then, we can write the following integral form for four-point functions in $SL(2,\R )_k$ WZNW model (see the analogous construction in \cite{fz4p} and \cite{a4p})
\begin{eqnarray*}
{\mathcal A}^{WZNW}_{j_{1},j_{2},j_{3},j_{4}}&=&\prod_{a<b}\left| x_{a}-x_{b}\right| 
^{2J_{ab}}\left| z_{a}-z_{b}\right|^{2h_{ab}} \left| 
z\right|^{-\frac {4j_{1}j_{2}}{k-2}} \left| 1-z\right|^{-\frac {4j_{1}j_{3}}{k-2} } 
\prod_{r=1}^{-2j_{1}} \times \nonumber \\
&&\times \int d^{2}w_{r}\prod_{r=1}^{-2j_{1}}\left| w_{r}\right| ^{\frac {2}{k-2}
(1-k-j_{1}-j_{2}+j_{3}+j_{4})} \left| 1-w_{r}\right| ^{\frac {2}{k-2} 
(1-k-j_{1}+j_{2}-j_{3}+j_{4})} \times
\end{eqnarray*}
\begin{eqnarray}
&&\times \left|
z-w_{r}\right| ^{\frac {2}{k-2} s}\left| x-w_{r}\right|
^{2} \prod_{r<t}^{-2j_{1}-1,-2j_{1}}\left| w_{r}-w_{t}\right| ^{-\frac {4}{k-2} } 
f(j_1,j_2,j_3,j_4;k)  \nonumber \\
&&  \label{p4puntos}
\end{eqnarray}
where $s=1-j_{1}-j_{2}-j_{3}-j_{4}$. And where the normalization factor $f(j_1,j_2,j_3,j_4;k)$ is specified by the particular normalization of two-point functions and the structure constants.

\subsubsection*{Normalization, reflection coefficient and structure constants}

In references \cite{fz4p} and \cite{a4p} the normalization of four-point functions was inferred from the normalization of reflection coefficients and, thus, the integral formula was used to derive the structure constants of the WZNW model.

Here, we fix the normalization by writing the appropriate\footnote{{\it cf.} \cite{fz4p,a4p}. The normalization presented here corresponds to the $\hat {sl(2)_k}$ structure constants.} function $f(j_1,j_2,j_3;k)$ which leads to the reflection coefficients $B(j)$ and the structure constants $C(j_1,j_2,j_3)$. This normalization factor is, then, given by
\begin{eqnarray}
f(j_1,j_2,j_3,j_4;k) &=& W(b) \left( \pi \frac {\Gamma (1-b^2)}{\Gamma (1+b^2)}\right) ^{1-\sum _{\rho =1}^4 j _{\rho }} \left( \pi \mu _+ \frac {\Gamma (1+b^2)}{\Gamma (1-b^2)} b^{-2b^2} \right) ^{b^{-1}(Q-\sum _{\nu =1}^4 \alpha _{\nu }-b^{-1}/2)} \times \nonumber \\
&& \times  \prod _{\mu = 1}^4 \frac {\Upsilon (-2j_{\mu}b-b)}{\Upsilon (2\alpha _{\mu})} \label{normaton}
\end{eqnarray}
where the $\Upsilon $ functions are given by
\[
\log \Upsilon (x)=\frac{1}{4}\int_{0}^{\infty }\frac{d\tau }{\tau }\left(
b+b^{-1}-2x\right) ^{2}e^{-\tau }-\int_{0}^{\infty }\frac{d\tau }{\tau }%
\frac{\sinh ^{2}\left( \frac{\tau }{4}(b+b^{-1}-2x)\right) }{\sinh \left( 
\frac{b\tau }{2}\right) \sinh \left( \frac{b^{-1}\tau }{2}\right) } 
\]
and have zeros in the lattice 
\begin{eqnarray}
x \in -b\Z_{\geq 0}-b^{-1}\Z_{\geq 0} \ , \ \ x \in b\Z_{>0}+b^{-1}\Z_{>0},  \label{lcdtm}
\end{eqnarray}
and where $W(b)$ is a $j_{\mu }$-independent numerical factor whose explicit form can be found in the literature \cite{teschner}. 

Thus, (\ref{normaton}) provides the complete integral representation (\ref{p4puntos}) for four-point functions in WZNW on $SL(2,\R)$. We shall study its properties in the following paragraphs. And let us begin by showing that, indeed, the normalization proposed here reproduces the two-point and three-point functions in the corresponding particular cases ${\mathcal A}^{WZNW}_{0,0,j_3,j_4}$ and  ${\mathcal A}^{WZNW}_{0,j_2,j_3,j_4}$ as claimed.

Indeed, replacing $j_{1}=j_{2}=0$ in (\ref{p4puntos}) we have
\begin{eqnarray}
{\mathcal A}^{WZNW}_{0,0,j_3,j_4}=\left| z_{3}-z_{4}\right| ^{-4h_{j_3}} \left( \left|
x_{3}-x_{4}\right| ^{-4j_3} B(j_3) \delta (j_3-j_4) + \delta ^{(2)} (x_3-x_4) \delta (j_3+j_4-1) \right)  \nonumber \\
\label{cacarulo1}
\end{eqnarray}
being
\begin{eqnarray}
B(j)= \frac {b^{-2}}{\pi } \left( \pi \frac{\Gamma \left( 1-b^2 \right) }{%
\Gamma \left( 1+b^2 \right) }\right) ^{1-2j} \frac {\Gamma ( 1+b^2(1-2j))}{\Gamma ( b^2 (2j-1))}
\end{eqnarray}
This $SL(2,\R )_k$ reflection coefficient satisfies
\begin{eqnarray}
B(j)B(\sigma (j)) &=& \frac {1}{\pi ^2} (2j -1)(2\sigma (j) -1) \\
B(j)B(s(j)) &=& \frac {1}{\pi ^2} B( 1/2b^2) \\
B(j) B(\tilde {\sigma} (j)) &=& \frac {1}{\pi ^2}  \frac {B^2 (1/2b^2)}{(2s(j)-1)(2s(\tilde {\sigma } (j)) -1)}
\end{eqnarray}
One of the most remarkable properties, among these, is the fact that $B(j)B(s(j)) = \frac {1}{\pi ^2 b^4} \left( \pi \frac {\Gamma (1+b^2)}{\Gamma (1-b^2)}\right)^{b^{-2}}$ does not depend on $j$ (see \cite{mo3}). And we also find $B(0) =-1$, $\lim _{j \rightarrow \frac 12} B(j) = 0$, $\lim _{b \rightarrow 1, \frac 12 <j} B(j) = 0$.

In a similar way, one finds the following expression for three-point functions for the case ${\mathcal A}^{WZNW}_{j_{1}=0,j_{2},j_{3},j_{4}}$, {\it i.e.}
\begin{eqnarray}
{\mathcal A}^{WZNW}_{0,j_{2},j_{3},j_{4}}=\prod_{a<b}\left| z_{a}-z_{b}\right|
^{2(h_c -h_a -h_b )}\left| x_{a}-x_{b}\right| ^{2(j_c -j_a -j_b )} C(j_{2},j_{3},j_{4})  \label{cacarulo2}
\end{eqnarray}
with $a,c,b \in \{2,3,4 \}$ and
\[
C(j_{2},j_{3},j_{4})= \frac {1}{2\pi ^3 b^2} \left( \pi \frac{\Gamma \left(1-b^2 \right) }{\Gamma \left( 1+b^2 \right) }\right) ^{2-j_2-j_3-j_4} \frac{G_k (1-j_{2}-j_{3}-j_{4})}{G_k (-1)} \prod _{a=2}^{4}   \frac {G_k (2j_a-j_{2}-j_{3}-j_{4})}{G_k (1-2j_{a})}
\]
In this expression, $G_k (x)$ are functions\footnote{The physical information of the formula is codified in the analytic properties of $G$ functions; for our purpose it is enough to mention that $G_k (x)$ presents poles in the set $x \in \Z _{<0}+ \Z _{<0} (k-2)$ and $x \in \Z _{\geq 0}+ \Z _{\geq 0} (k-2)$.} which can be written in terms of $\Gamma _2$ Barnes functions, namely
\[
G_{k}(x)\equiv (k-2)^{\frac{x(k-1-x)}{2k-4}}\Gamma _{2}(-x\mid 1,k-2)\Gamma _{2}(k-1+x\mid 1,k-2) 
\]
where 
\[
\log (\Gamma _{2}(x\mid 1,y))\equiv \lim_{\varepsilon \rightarrow 0}\frac {\partial}{
\partial \varepsilon } \left( \sum_{n=0}^{\infty }\sum_{m=0}^{\infty
}(x+n+my)^{-\varepsilon } + \sum_{n=0}^{\infty }\sum_{m=0}^{\infty
}(\delta _{n,0}\delta _{m,0}-1) (n+my)^{-\varepsilon }\right)
\]
And, alternatively, these can be written in terms of the $\Upsilon _b $ functions introduced in Liouville literature by Zamolodchikov and Zamolodchikov, see for instance \cite{mo3,teschner,kutasov,nos3} and references therein.

The structure constants are invariant under any permutation of the set $\{ j_{2},j_{3},j_{4} \}$ and satisfy the following property under Weyl reflection $\sigma$
\[
C(j_{2},j_{3},j_{4})=N(j_{2},j_{3},j_{4}) C(j_{2},\sigma (j_{3}),j_{4})
\]
being
\[
N(j_{2},j_{3},j_{4})= \frac {1}{\pi } B^{-1} (j_3) \frac {\Gamma (1+j_2-j_3-j_4)\Gamma (1-j_2-j_3+j_4)\Gamma (2j_3)}{\Gamma (-j_2+j_3+j_4)\Gamma (j_2+j_3-j_4)\Gamma (1-2j_3)}
\]
A similar relation, for certain\footnote{clearly, these satisfy $N(j_{2},j_{3},j_{4})N(j_{2},\sigma (j_{3}),j_{4})= \tilde {N} (j_{2},j_{3},j_{4}) \tilde {N} (j_{2},\tilde {\sigma } (j_{3}),j_{4})=1$, and permutations.} $\tilde {N} (j_{2},j_{3},j_{4})$, can be obtained for $C(j_{2},\tilde {\sigma } (j_{3}),j_{4})$ by using the functional properties of these special functions. We also find the following (see \cite{mo3,becker,nos3})
\begin{eqnarray}
C(j,j,0) &=& B(j) \delta (0) \\
C(j,\sigma (j),\sigma (0)) &=& \frac {1}{\pi } (2j -1)(2\sigma (j) -1) \delta (0) \\
C(j,s(j),s(0)) &=& \frac {1}{\pi } B( 1/2b^2) \delta (0)
\end{eqnarray}
which realize the relation between the structure constants and reflection coefficients; namely $C(j,\sigma (j), \sigma (0))=\pi B(j)B(\sigma(j))\delta (0)$ and $C(j,s (j), s (0))=\pi B(j)B( s(j))\delta (0)$.

Let us observe that (\ref{cacarulo1}) and (\ref{cacarulo2}) manifestly show the CFT$_2$ structure of the worldsheet theory in terms of ($z,\bar z$) and the dual theory in terms of ($x,\bar x$).

\subsubsection*{Particular examples and monodromy invariance}

The general form of the integral representation for the four-point functions is highly non trivial, but it can be explicitly solved for several particular cases; these were analysed in \cite{ponsot,fz4p}. Let us consider here the example $n_-=0$, $n_+=1$. In this particular case, one finds that the four-point function is proportional to the following integral 
\begin{eqnarray*}
\left| I_{k}(j_2,j_3,j_4;z,x)\right| ^2&=&\int d^{2}w \left| w\right| ^{\frac {2}{2-k} (-j_{2}+j_{3}+j_{4}+\frac 32-k)}\left|
1-w\right| ^{\frac {2}{2-k} (-j_{2}+j_{3}-j_{4}-\frac 32+k)} \times \\
&&\times\left| z-w\right| ^{\frac {2}{k-2}(-j_{2}-j_{3}-j_{4}+\frac 32)}\left| x-w\right| ^{2}
\end{eqnarray*}
whose solution is polynomial in $x$ and $\bar x$; namely \cite{fz4p}  
\begin{eqnarray*}
\left| I_{k}(j_{2},j_{3},j_{4};z,x)\right| ^2= X_{0,0} (z,\bar z)+X_{0,1} (z,\bar z) x+\bar {X} _{0,1} (z,\bar z) \bar x +X_{1,1} (z,\bar z) |x|^2
\end{eqnarray*}
where the functions $X_{i,j}$ can be expressed in terms of hypergeometric functions for $(z,\bar z)$.

This provides the explicit result for the particular configuration $j_{1}=\frac{1}{2}$ and generic values of  $j_{2},j_{3},j_{4}$ \cite{ponsot,fz4p}. And, it is feasible to show that different terms in the power expansion of $(x,\bar x)$ are monodromy invariant in $(z, \bar z)$. For instance, let us write the first term of the polynomial in the following convenient way
\begin{eqnarray*}
X_{0,0}(z,\bar{z}) &=&\frac{\Gamma (1-\tilde \gamma) \Gamma (\tilde \alpha) \Gamma (\tilde \gamma -\tilde \alpha)}{\Gamma (\tilde \gamma) \Gamma (1-\tilde \alpha) \Gamma (1- \tilde \gamma +\tilde \alpha )}\left( \left| F(\tilde \alpha ,\tilde \beta ,\tilde \gamma ,z)\right| ^{2}+\tilde \lambda \left| z^{1-\tilde \gamma }F(\tilde \alpha -\tilde \gamma +1,\tilde \beta -\tilde \gamma +1,2-\tilde \gamma ,z)\right|^{2}\right)
\end{eqnarray*}
where $ \tilde \alpha = \frac {1}{2-k} (1/2-j_2-j_3+j_4) \ , \ \tilde \beta = \frac {1}{2-k} (3/2-j_2-j_3-j_4) \ , \ \tilde \gamma = \frac {1}{2-k} (1-2j_2) $ and
\begin{eqnarray*}
\tilde \lambda = -\frac{\Gamma (\tilde \alpha -\tilde \gamma +1)\Gamma (\tilde \beta -\tilde \gamma +1)\Gamma^{2}(\tilde \gamma ) \Gamma (1-\tilde \alpha) \Gamma (1-\tilde \beta)}{\Gamma (\tilde \alpha )\Gamma (\tilde \beta ) \Gamma (\tilde \gamma -\tilde \alpha) \Gamma (\tilde \gamma -\tilde \beta) \Gamma ^2 (2-\tilde \gamma)}
\end{eqnarray*}
which, in fact, we recognize as a monodromy invariant linear combination.

Analogously, we can analyse other terms of the expansion; $e.g.$ we can write
\begin{eqnarray*}
X_{1,1}(z,\bar{z}) &=&\frac{\Gamma (1-\hat \gamma) \Gamma (\hat \alpha) \Gamma (\hat \gamma -\hat \alpha)}{ \Gamma (\hat \gamma) \Gamma (1-\hat \alpha) \Gamma (1- \hat \gamma +\hat \alpha )}\left( \left| F(\hat \alpha ,\hat \beta ,\hat \gamma ,z)\right| ^{2}+\hat \lambda \left| z^{1-\hat \gamma }F(\hat \alpha -\hat \gamma +1,\hat \beta -\hat \gamma +1,2-\hat \gamma ,z)\right|^{2}\right)
\end{eqnarray*}
being $\hat \alpha = 1 + \tilde \alpha $, $\hat \beta = \tilde \beta $, $\hat \gamma = 1 + \tilde \gamma $ and $ \hat {\lambda} (\hat {\alpha }, \hat {\beta },\hat {\gamma })= \tilde {\lambda} (1+\tilde {\alpha }, \tilde {\beta },1+\tilde {\gamma })$. And, again, this turns out to be a monodromy invariant function.

At this point, one could naively suppose that since these are monodromy invariant solutions of a certain hypergeometric equation, these vanish in the particular point where the indicial exponent becomes zero because of a cancellation of both terms. But it is not the case due to the particular overall factors standing in $X_{0,0}$ and $X_{1,1}$. In fact, if the limit is taken adequately, a finite remnant contribution appears containing logarithmic terms $\sim \log|z|$ in the point $ j_2=\frac 12$ for the term $X_{0,0}(z,\bar z)$ and in the point $ j_2= \frac {k-1}{2}$ for the term $X_{1,1}(z,\bar z)$. We also find logarithmic contributions\footnote{ and, of course, similarly for $j_4 = \frac 12$.} $\sim \log (1-z)$ for $j_3 = \frac 12$. These logarithmic solutions are different from those analysed previously; these present logarithmic terms in the world-sheet variables $(z,\bar z)$ and are polynomial in $(x,\bar x)$. These solutions for $j_1 =\frac 12$ are independent of the outgoing momenta $j_3,j_4$. In order to try a classification of singularities and propose the corresponding interpretation it is necessary to discriminate between the ones which correponds to physical processes and those which do not. In fact, let us notice that the solutions mentioned in this subsection are unphysical solutions because these are excluded by the unitarity bound (\ref{izabel}).

In some aspects, this is similar to the cases of the logarithmic solutions standing in the fixed points of Weyl transformations. 

Hence, we see how the unitarity bound \cite{mo1} on the free spectrum and the locality constraints \cite{mo3} on $N$-point functions exclude the $z$-dependent logarithmic terms in four-point correlators.

\subsubsection*{More on pole structure at $z=x$. }

In a different context, and with the intention to emphasize the power of such a description of the four-point function, let us analyse a second particular case by considering $n_-=1$ and $n_+=0$, which clearly implies $j_1=1-\frac k2$. For this configuration, the four-point function would be proportional to the following integral
\begin{eqnarray*}
\left| J_{k}(j_2,j_3,j_4;z,x)\right| ^2&=&\int d^{2}v \left| v\right| ^{2(-\frac k2-j_{2}+j_{3}+j_{4})}\left|
1-v\right| ^{2(-j_{2}+j_{3}-j_{4}-\frac k2)}\times  \\
&&\times \left| z-v\right| ^{2(\frac k2-j_{2}-j_{3}-j_{4})}\left| x-v\right| ^{2(k-2)}
\end{eqnarray*}
From this, by rescaling the variables, it is possible to see again the arising of the singular factor $(x-z)^{k-j_1-j_2-j_3-j_4} $ in the limit $x \rightarrow z$.

Then, in this subsection we have made use of particular examples to study different aspects of the four-point functions: we obtained the normalization factor by means of the requirement that the $j_1=0$ case corresponds to the structure constants, we observed the monodromy invariance of the case $j_1=-\frac 12$ and we obtained the singular behavior (\ref{i}) by studying the case $j_1=1-\frac k2$. In \cite{mo3}, another particular case ($j= \frac k2$) satisying a BPZ-like equation was also studied to obtain this singular behavior at $x=z$.

\subsubsection*{On $\hat {sl(2)} _k$ admissible representations}

In reference \cite{ay} the fusion rules for $\hat {sl(2)} _k$ admissible representations were explicitly written down. Andreev made use of this adaptation of the Fateev-Zamolodchikov equivalence between minimal models and WZNW theory in order to calculate the structure constants and reobtain in that way the fusion rules of $SL(2,\R)_k$ admissible representations \cite{a4p,ay}.

These representations are constructed by the following particular configurations of $j$ and imposing the restriction of $k$ being a rational number, namely
\begin{equation}
j \in \frac {1+ \Z _{\leq 0}}{2} + \frac {1+ \Z _{\leq 0}}{2} (k-2) \ , \ \ \ j \in \frac {\Z _{> 0}}{2} + \frac {\Z _{\geq 0}}{2} (k-2)   \label{adm}
\end{equation}
In \cite{a4p} the respective integral representations were given for both series in (\ref{adm}) by explicitly discerning among them, and by noticing that the values corresponding to the first case (the $l.h.s.$) includes the identity operator $\Phi _{j =0}$. This fact enabled the author to fix the normalization by considering the case $j_1=0$ and, thus, to obtain the structure constants which leads to fusion rules.

Let us notice that, similarly, the second case includes $\Phi _{j=\frac k2 }$ which corresponds to the conjugate representation of the identity operator; consequently, it also leads to a simple expression ({\it i.e.} without integrals for $j_1 = \frac k2$) and the structure constants are uniquely determined by the normalization of the mentioned conjugate identity.

\subsubsection*{Factorization limit of the integral expression}

A detailed analysis of the factorization limit of the four-point function in string theory on $AdS_3$ has been recently performed in reference \cite{mo3} by Maldacena and Ooguri. Here, let us examine the factorization properties of the integral formula (\ref{p4puntos}) in the limit $z\rightarrow 0$.

Starting with the integral expression (\ref{p4puntos}), separating $l$ of the $-2j_{1}$ integrals and defining the change of variables $y_{i\leq l} \equiv z^{-1}w_{i\leq l}$ while $w_{i>l}$ and $x$ remain invariant, we find the following schematic behavior in the region $z \approx 0$
\begin{equation}
{\mathcal A}^{WZNW}_{j_{1},j_{2},j_{3},j_{4}} \sim \left|
z \right| ^{2p}\left| x\right|^{2l}\times \left( 1+{\mathcal O} \left( |z|^2 , \partial \bar {\partial }   \right)  \right) \nonumber
\end{equation}
where $p$ is given by
\begin{eqnarray*}
p = - \frac{2j_{1}j_{2}+2(j_{1}+j_{2})l+l(l-1)}{k-2}
\end{eqnarray*}
being $l\in \N$ covering the range $[0,-2j_1]$. By using the fact that $h_{j_1}=h_{j_2}=1$ and by integrating in $d^2z$ we finally obtain pole conditions in $p+1 \in \Z_{\leq 0}$, which we can rewrite as follows
\begin{eqnarray*}
-\frac{j(j-1)}{k-2}+n=1 \ , \ \ \ j=j_{1}+j_{2}+l  \label{xcvc}
\end{eqnarray*}
where $n \in \N $ and, again, where $l$ is an integer number such that\footnote{%
This is due to the fact that $l$ is precisely the amount of screening charges selected to be taken together with $z$ in the coincidence limit.}
\begin{equation}
j_{2}-j_{1} \leq j\leq j_{2}+j_{1} \label{vi23}
\end{equation}
It is easy to perform the analogous limit when the realization includes both types of screening charges; in this case we would obtain $j=j_{1}+j_{2}+l+l^{\prime }(k-2)$ which again implies $0<l+l^{\prime } (k-2)<-2j_{1}$. Hence, we show in this way that the pole conditions appearing in the coincident limit $z\rightarrow 0$ agree with the mass-shell conditions for string states characterized by a quantum number $j$ defined in (\ref{xcvc}). Moreover, the four-point function has the following functional form in the mentioned limit ${\mathcal A}^{WZNW}_{j_{1},j_{2},j_{3},j_{4}}\sim \left| z\right|^{2(h_{j}-h_{j_{1}}-h_{j_{2}})}\left| x\right|^{2(j-j_{1}-j_{2})}$.

\section{Appendix B: Liouville theory description of symmetries of KZ equation}

It deserves a separate appendix to mention another interesting point about this correspondence between WZNW model and Liouville CFT. Indeed, it is feasible to use this correspondence in order to study certain symmetries of the correlators of both models (of both {\it sides}). To be more precise, for a given five-point function in Lioville theory with primary fields characterized by vertex operators $e^{2\alpha _i \varphi }$ being $\{ \alpha _1,\alpha _2,\alpha _3,\alpha _5= -\frac {1}{2b},\alpha _4 \}$, it is possible to see that the changes\footnote{Notice that such changes represent a symmetry of Liouville theory since states with $\alpha$ and states with $\tilde \alpha = b+b^{-1}-\alpha $ have the same conformal dimension.} $\alpha _{1,2,3,4} \rightarrow Q-\alpha _{1,2,3,4 } $ induce the following changes in the quantum numbers of $SL(2,\R)$-states: $j_{1,2,3,4 } \rightarrow \frac{k}{2}-j_{1,2,3,4 }$. Reciprocally, if the transformations $j_{2,3} \rightarrow \frac{k}{2}-j_{2,3} $ are performed, leaving invariant $j_1$ and $j_4$, then the following corresponding changes in the {\it Liouville CFT side} are found: $\alpha _{1} \leftrightarrow \alpha _{4}$ and $\alpha _{2} \leftrightarrow \alpha _{3}$. These are some examples of the map which we summarize below in more detail.

Now, we can explore how to use the dictionary in order to map the reflection transformations in Liouville CFT into non trivial transformations among the quantum numbers of $SL(2,\R )_k$ representations. One could naively expect that the change on the $SL(2,\R )_k$ {\it side} induced by Liouville reflection could correspond to Weyl reflections $\sigma$, $\tilde \sigma$ or combinations thereof, but this is not the case. Indeed, it is possible to see that the corresponding transformations in the $SL(2,\R )_k$ representation indices turn out to be a non trivial combination of spectral flow automorphism and $r$-symmetry. Let us show it explicitly.

We begin with the simplest cases of $\Z _2$ trasformations performed on the {\it Liouville side}. The following conventions and nomenclature are used in the list below: the greek indices $\mu$,$\nu $ cover $\{1,2,3,4\}$ while the latin indices $i,j,k$ cover only $\{2,3,4\}$; besides, the latin indices $i,j,k$ have to be considered distinct in each sentence even though similar expressions will hold for any permutation of the set $\{i,j,k \}$; the omission of any index $j_{\mu }$ or $\alpha _{\mu }$ implies, in our convention, that it remains invariant. Besides, if certain sentence includes a greek index (let us say $\mu $) then it has to be interpreted as the transformation of {\it all} the quantum numbers ({\it i.e.} not a generic one).

\subsubsection*{$\Z _2$ symmetries and crossing relations}
The simplest nilpotent transformations are defined by interchanges among certain pairs of indices in the correlators; for instance, the following correspondence holds\footnote{{\it i.e.} the transformation for Liouville states in the {\it l.h.s.} corresponds to the transformation of $SL(2,\R)_k$ in the {\it r.h.s.} and {\it viceversa}.}
\begin{align}
\alpha _j \leftrightarrow Q- \alpha _k \ \ \ \ corresponds \ to \ \ \ \ j_1 \leftrightarrow j_i \label{n1}
\end{align} 
where it is necessary to take into account that our convention implies that $i \neq j \neq k \neq i$.

Taking into account the distinctive role played by the first indices $j_{\mu =1}$ and $\alpha _{\mu =1}$, we can consider a similar but independent transformation\begin{align}
\alpha _i \leftrightarrow \alpha _j \ \ \ \ corresponds \ to \ \ \ \ j_i \leftrightarrow j_j\label{n2}
\end{align} 
These trivial relations follow directly from the definition of $\alpha _{\mu}$'s.
\subsubsection*{$\Z _2$ symmetries and spectral flow}
A less trivial relation is represented by the following correspondence
\begin{align}
\alpha _{\mu } \leftrightarrow Q- \alpha _{\mu } \ \ \ \ corresponds \ to \ \ \ \ j_{\mu } \leftrightarrow s(j_{\mu })  \label{n3}
\end{align} 
where, again, the appearance of the subindex $\mu $ in this sentence means that {\it all} the quantum numbers are interchanged ({\it i.e.} not a generic one). Here, the relation between the reflection symmetry in Liouville theory and the spectral flow symmetry in the $\hat {sl(2)}_k$ algebra is manifest. Besides, other relations realizing this aspect are given by
\begin{align}
\left. \begin{array}{r} \alpha _1 \leftrightarrow \alpha _k \\ \alpha _j \leftrightarrow \alpha _i  \end{array} \right\} \ \ corresponds \ to \ \ \left\{ \begin{array}{l} j_i \leftrightarrow s(j_i) \\ j_j \leftrightarrow s(j_j) \end{array} \right.  \label{n4}
\end{align} 
and similarly by
\begin{align}
\alpha _1 \leftrightarrow \alpha _k \ \ \ \ corresponds \ to \ \ \ \ j_i \leftrightarrow s(j_j)\label{n5}
\end{align} 
The automorphism $s$ commutes {\it standard} ({\it i.e.} non-flowed) discrete representations.

Notice that (\ref{n4}), once combined with equation (\ref{parcialito}), leads to a relation between the logarithmic solutions located at the points $j_1+j_i=j_j+j_k$ and $k=j_1+j_2+j_3+j_4$ of the momenta space since, for instance, the interchange of quantum numbers $\alpha _1 \leftrightarrow \alpha_2$ corresponds to the change $j_1+j_2-j_3-j_4 \leftrightarrow j_1+j_2+j_3+j_4-k $. Notice also that (\ref{entreomegas}), which establishes a relation between the leading terms of both solutions ${\mathcal F}_{j_1,j_2,j_3,j_4}(z;x)$ and ${\mathcal F}_{j_1,s (j_2),s (j_3),j_4}(z;zx^{-1})$, is implied by the symmetry under the interchange $\alpha _1 \leftrightarrow \alpha_3$ in Liouville correlators ({\it i.e.} which, consistently, corresponds to the replacement of inserting points $z \leftrightarrow 1$). These observations manifestly show the usefulness of this dictionary for the study of symmetries of the KZ equation.
\subsubsection*{$\Z _2$ symmetries and $r$-transformations}
The following $\Z _2$ transformations permits to observe how the reflection in Liouville correlators induces non-trivial transformations in the WZNW theory; namely
\begin{align}
\alpha _i \leftrightarrow Q- \alpha _i \ \ \ \ corresponds \ to \ \ \ \ \left\{ \begin{array}{l} j_i \leftrightarrow \frac 12 \sum _{\mu =1}^{4} j_{\mu } - j_1 \\ j_j \leftrightarrow \frac 12 \sum _{\mu =1}^{4} j_{\mu } - j_k \end{array} \right.  \label{n13}
\end{align} 
which includes the $r$-transformation as the particular case $\{i,j,k\} = \{3,4,2\}$, {\it i.e.} $r$-transformation corresponds to transforming $\alpha _3 \rightarrow Q- \alpha _3$ in Liouville five-point correlators, namely 
\begin{align}
\alpha _3 \leftrightarrow Q- \alpha _3 \ \ \ \ corresponds \ to \ \ \ \ & j_{\mu} \leftrightarrow r(j_{\mu})
\end{align} 
These symmetry transformations were catalogued under the titles of $strange$ \cite{nichols} and $impressive$ \cite{a4p} $\Z _2$ {\it symmetries} in the literature, where it was signaled that these correspond to symmetries because of the fact that both ${\mathcal F}_{j_1,j_2,j_3,j_4}(z;x)$ and $z^{-h_{r(j_1)}-h_{r(j_2)}}(1-z)^{-h_{r(j_3)}-h_{r(j_4)}} {\mathcal F}_{r(j_1),r(j_2),r(j_3),r(j_4)}(z;x) $ satisfy the KZ equation. Here, we derived them in a very natural way from the reflection relations in Liouville CFT description.

On the other hand, we also find 
\begin{align}
\alpha _1 \leftrightarrow Q- \alpha _1 \ \ \ \ corresponds \ to \ \ \ \ j_{\mu } \leftrightarrow \frac 12 \sum _{\nu = 1}^{4} s(j_{\nu }) - s(j_{\mu }) \label{n14}
\end{align} 
We can also consider $\Z _3$ transformations; namely

\subsubsection*{$\Z _3$  symmetries}
First, let us consider
\begin{align}
\alpha _i \rightarrow \alpha _j \rightarrow \alpha _k \rightarrow \alpha _i \ \ \ \ corresponds \ to \ \ \ \ j_k \rightarrow j _i \rightarrow j _j \rightarrow j _k \label{n6}
\end{align} 
Similarly, we have the following
\begin{align}
\alpha _i \rightarrow Q- \alpha _k \rightarrow \alpha _j \rightarrow \alpha _i \ \ \ \ corresponds \ to \ \ \ \ j_1 \rightarrow j _i \rightarrow j _j \rightarrow j _1 \label{n7}
\end{align} 
and certain reciprocal version of this
\begin{align}
\alpha _1 \rightarrow \alpha _i \rightarrow \alpha _j \rightarrow \alpha _1 \ \ \ \ corresponds \ to \ \ \ \ j_i \rightarrow s(j _k) \rightarrow j _j \rightarrow j _i \label{n12}
\end{align} 
On the other hand, we can also consider other transformations.
\subsubsection*{$\Z _4$ symmetries}
For instance, we have
\begin{align}
\alpha _1 \rightarrow \alpha _i \rightarrow \alpha _j \rightarrow \alpha _k \rightarrow \alpha _1 \ \ \ \ corresponds \ to \ \ \ \ \left\{ \begin{array}{l} j_i \rightarrow s(j _k) \rightarrow s(j _i) \rightarrow j _k \rightarrow j _i \\ j_j \rightarrow s(j _j) \rightarrow j _j \rightarrow s(j _j) \rightarrow j _j \end{array} \right.  \label{n10}
\end{align} 
which corresponds to a $\Z _4$ transformation decomposable on the right hand side as $\Z _4 \oplus \Z _2$; and
\begin{align}
\left. \begin{array}{l} \alpha _j \rightarrow Q- \alpha _j \rightarrow \alpha _j \rightarrow Q-\alpha _j \rightarrow \alpha _j \\  \alpha _i \rightarrow Q- \alpha _k \rightarrow Q- \alpha _i \rightarrow \alpha _k \rightarrow \alpha _i  \end{array} \right\} \ \ \ \ corresponds \ to \ \ \ \ j_ 1 \rightarrow j _i \rightarrow j _j \rightarrow j _k \rightarrow j _1  \label{n8}
\end{align} 
which, again, can be decomposed on one side as $\Z _4 \oplus \Z _2$.

\subsubsection*{On scattering of winding strings in $AdS_3$.}

Notice that, as it occurs with (\ref{n7}) and (\ref{n12}), each condition listed above includes a reciprocal version which corresponds to interchanging the roles played by $j_{\mu } \leftrightarrow \alpha _{\mu }$ and $\tilde \alpha _{\mu } = Q-\alpha _{\mu } \leftrightarrow s(j_{\mu })$. For instance, (\ref{n1}) is the reciprocal to (\ref{n5}), (\ref{n10}) is the reciprocal to (\ref{n8}), and (\ref{n3}) is its own reciprocal.

Then, by considering the reciprocal functorial relations of (\ref{n13}) and (\ref{n14}), this observation would lead us to obtain the four-point correlators including one index that, instead $j_{\nu }$, has been replaced by $\frac k2- j_{\nu }$. Moreover, taking into account the identification holding for states in the sectors $\omega =0 $ and $\omega =1$ of spectral {\it flowed} representations, perhaps it could be possible to interpret such correlation functions as describing processes involving one string state of momentum $j_{\nu }$ and winding $\omega _{\nu } =1$ in $AdS_3$. This seems to be consistent with the analogous results for three-point functions analysed in \cite{mo3}. Besides, since in that work it was shown how the three-point function representing scattering processes violating winding number in $AdS_3$ can be written in terms of the analogous quantities for the conservative case, we can write here that precise factor which relates the structure constants for both cases as
\begin{eqnarray}
\frac {B(1+b^{-2}-b^{-1}(\alpha _1+\alpha _i)) \Gamma (-b^{-2}+2b^{-1} \alpha _i)}{ \Gamma (1+b^{-2}-2b^{-1} \alpha _i)}
\end{eqnarray}
which is the quotient between the structure constants for winding violating and conserving processes involving three string states characterized by quantum numbers $(j_2,j_3,j_4)$ and $(\omega _i =1, \omega _{j,k} =0)$ (and consequently $j_1 =0, \omega _1 =0$). In the above factor it is necessary to consider $2\alpha _1 = b(j_2+j_3+j_4-2j_i+\frac {b^{-2}}{2})$, $2\alpha _i = b(-j_2-j_3-j_4+2+\frac 32 b^{-2})$ and $2\alpha _{j,k} = b(j_2+j_3+j_4-2j_{k,j}+\frac {b^{-2}}{2})$, which correspond to the replacement $j_1 \rightarrow 0 , j_i \rightarrow s(j_i ), j_{j,k} \rightarrow j_{j,k}$.

In the following subsection, we answer the question about how to see the appearance of logarithmic solutions in terms of the five-point correlators involving Liouville primary fields.

\section{Appendix C: Explicit solutions for the case $k-2j \in \Z$}

Let us present in this appendix some explicit expression for $j$-dependent logarithmic solutions.

Standard formulae concerning hypergeometric equations lead to the following form for the general solutions: 
\subsubsection*{The case $k-2j\in \Z_{>0}$} 
In this case, we have 
\begin{eqnarray*}
\tilde{{\mathcal G}}_{j_{1},j_{2,}j,j_{3},j_{4}}^{(0)}&=&F\left( j_1+j_2-j,j_3+j_4-j,k-2j,zx^{-1} \right) (1+\lambda (\log z -\log x+ \\
&&+\sum_{r=1}^\infty \left( zx^{-1} \right)^r\frac{\Gamma (k-2j)\Gamma (j_1+j_2-j+r)\Gamma
(j_3+j_4-j+r)}{\Gamma (k-2j+r)\Gamma (j_1+j_2-j)\Gamma (j_3+j_4-j)} \times\\
&&\times (\psi (j_1+j_2-j+r)+\psi (j_3+j_4-j+r)-\psi (j_1+j_2-j)+ \\
&&-\psi (j_3+j_4-j)-\psi (r+1)+\psi (1)+\psi (k-2j+1)-\psi (k-2j+1+r))+ \\
&&-\sum_{r=1}^{k-2j-1} \left( xz^{-1} \right)^{r}\frac{\Gamma (r)\Gamma (2j+1-k+r)\Gamma
(1+j-j_1-j_2)\Gamma (1+j-j_3-j_4)}{\Gamma (2j+1-k)\Gamma
(1+j-j_1-j_2+r)\Gamma (1+j-j_3-j_4+r)}))
\end{eqnarray*}
Notice that some momenta configurations leading to these solutions can occur within the range of unitarity (\ref{izabel}). The point $j=\frac {k-1}{2}$ is also included. In this particular case, the reflection symmetry connecting both terms in the monodromy invariant solution breaks down.
\subsubsection*{The case $k-2j\in \Z_{\leq 0}$} 
This configurations lead to consider the following basis of solutions
\[
\tilde{{\mathcal G}}_{j_{1},j_{2,}j,j_{3},j_{4}}^{(0)}=\left( zx^{-1} \right)^{1+2j-k}F\left( j+j_1+j_2+1-k,j+j_3+j_4+1-k,2j+2-k,zx^{-1} \right)
(1+\lambda (\log zx^{-1} +
\]
\begin{eqnarray*}
&&+\sum_{r=1}^\infty \left( zx^{-1} \right)^{1+2j-k+r}\frac{\Gamma (2j-k+2)\Gamma
(j_1+j_2+j+1-k+r)\Gamma (j_3+j_4+j+1-k+r)}{\Gamma (k+1)\Gamma (2j-k+2+r)\Gamma
(j_1+j_2+j-k)\Gamma (j_3+j_4+j+1-k)} \\
&&(\psi (j_1+j_2+j+1-k+r)+\psi (j+j_3+j_4+1-k+r)-\psi (2+2j-k+r)+ \\
&&-\psi (r+1)-\psi (j+j_1+j_2+1-k)-\psi (j+j_3+j_4+1-k)+\psi (2+2j-k)+\psi
(1))+ \\
&&-\sum_{r=1}^\infty \left( zx^{-1} \right)^{1+2j-k-r}\frac{\Gamma (r)\Gamma (k-2j-1+r)\Gamma
(1+k-j-j_1-j_2)\Gamma (1+k-j-j_3-j_4)}{\Gamma (k-2j-1)\Gamma
(1+k-j-j_1-j_2+r)\Gamma (1+k-j-j_3-j_4+r)}))
\end{eqnarray*}
These cases are excluded in the applications to string theory because of the unitarity bound imposed on the free spectrum and extra constraints on $N$-point functions.

And analogous solutions are obtained by replacing $j_1 \rightarrow s(j_1),$ $j \rightarrow s(j) $ and $j_4 \rightarrow s(j_4)$ for the cases $2j \in \Z$.

\section{Appendix D: On the normalization factor}

A normalization factor for the integral representation discussed in section 3 can be written as follows
\begin{eqnarray}
f(j_1,j_2,j_3,j_4;k) &=& \frac {1}{\pi b^2} \left( \pi \frac{\Gamma \left( 1- b^2\right) 
}{\Gamma \left(
1+ b^2 \right) }\right) ^{s}\left( \frac {\Gamma (-b^2)}{\Gamma (1+b^2)} \right)
^{2-2j_{1}}\prod_{a =1}^{4}\frac {\Gamma \left( 1+b^2 (1-2j_{a}) \right)}
{\Gamma \left( b^2 (2j_{a}-1) \right)}\prod_{r=1}^{s}\frac {\Gamma (-rb^2)}{\Gamma (1+rb^2)} 
\times
\nonumber \\
&&\times \prod _{a=2}^4 \prod_{r=1}^{j_{1}-j_{2}-j_{3}-j_{4}+2j_a} \frac {\Gamma (-rb^2)}{\Gamma (1+rb^2)}  \prod _{c=1}^4 \prod _{l=1}^{-2j_c} \frac {\Gamma (1+lb^2)}{\Gamma (-lb^2)} \nonumber \\
&&  \label{normaton2}
\end{eqnarray}
Certainly, by way of considerations whose details we will suppress, we proved that an extensive calculation which involves analytic continuation of $\Gamma $-functions formulae leads one to also find the following expression for three-point functions for the case ${\mathcal A}^{WZNW}_{j_{1}=0,j_{2},j_{3},j_{4}}$, {\it i.e.}
\begin{eqnarray}
{\mathcal A}^{WZNW}_{0,j_{2},j_{3},j_{4}}=\prod_{a<b}\left| z_{a}-z_{b}\right|
^{2(h_c -h_a -h_b )}\left| x_{a}-x_{b}\right| ^{2(j_c -j_a -j_b )} C(j_{2},j_{3},j_{4})  \label{cacarulo222}
\end{eqnarray}
up to a divergent factor $\sim \lim _{\epsilon \rightarrow 0} 2\pi ^2 \Gamma ^{-1} (\epsilon)$ which requires an appropriate regularization in the analytic continuation of the quotients of $\Gamma$ functions. The details are given in \cite{mitesisdoc}.



\newpage

\begin{thebibliography}{10}



\bibitem{ponsot}  B. Ponsot, Nucl.Phys. {\bf B642} (2002) 114-138.

\bibitem{rass}  J.L. Peterssen, J. Rasmussen and M. Yu, Nucl.Phys. {\bf B481} (1996) 577.

\bibitem{petko}  P. Furlan, A. Ganchev and V. Petkova, Nucl. Phys. {\bf B491} (1997) 635.

\bibitem{mo3}  J.M. Maldacena and Y. Ooguri, Phys.Rev. {\bf D65} (2002) 106006.

\bibitem{teschner}  J. Teschner, Nucl. Phys. {\bf B571} (2000) 555. J. Teschner, Nucl.Phys. {\bf B546} (1999) 369. J. Teschner, Nucl. Phys. {\bf B546} (1999) 390. J. Teschner, Phys.Lett. {\bf B521} (2001) 127-132. 

\bibitem{mo2}  J.M. Maldacena, Y. Ooguri and J. Son, J.Math.Phys. {\bf 42} (2001) 2961.
       
\bibitem{kz} V.G. Knizhnik and A.B. Zamolodchikov, Nucl.Phys. {\bf B247} (1984) 83.

\bibitem{log} A. Bilial and I.I. Kogan, Nucl.Phys. {\bf B449} (1995) 569. I.I. Kogan and A.M. Tsvelik, Mod.Phys.Lett. {\bf A15} (2000) 931. M. Gaberdiel, Int.J.Mod.Phys. {\bf A18} (2003) 4593. M. Flohr, Int.J.Mod.Phys. {\bf A18} (2003) 4497. A. Nichols, JHEP {\bf 0308} (2003) 040. J. Fjelstad, J. Fuchs, S. Hwang, A.M. Semikhatov and I.Y. Tipunin, Nucl.Phys. {\bf B633} (2002) 379. A. Bredthauer and M. Flohr, Nucl.Phys. {\bf B639} (2002) 450. I.I. Kogan and A. Lewis, Nucl.Phys. {\bf B509} (1998) 687. M.R. Gaberdiel and H.G. Kausch, Phys.Lett. {\bf B386} (1996) 131. M.R. Gaberdiel and H.G. Kausch, Nucl.Phys. {\bf B538} (1999) 631. Z. Maassarani and D. Serban, Nucl.Phys. {\bf B489} (1997) 603. M.A.I. Flohr, Int.J.Mod.Phys. {\bf A11} (1996) 4147. J.S. Caux, I.I. Kogan and A.M. Tsvelik, Nucl.Phys. {\bf B466} (1996) 444. Sanjay, J.Phys. {\bf A35} (2002) 9881. J. Rasmussen, {\it Applications of Free Fields in 2D Current Algebra }, arXiv:hep-th/9610167.

\bibitem{gurarie} V. Gurarie, Nucl.Phys. {\bf B441} (1995) 569.

\bibitem{liu}  H. Liu, Phys.Rev. {\bf D60} (1999) 106005.

\bibitem{ras1}  D.Z. Freedman, S.D. Mathur, A. Matusis, L. Rastelli, Phys.Lett. {\bf B452} (1999) 61.

\bibitem{ras2}  E. D'Hoker, D.Z. Freedman, S.D. Mathur, A. Matusis and L. Rastelli, Nucl.Phys. {\bf B562} (1999) 353.

\bibitem{ras3}  D.Z. Freedman, S.D. Mathur, A. Matusis and L. Rastelli. Nucl.Phys. {\bf B546} (1999) 96.

\bibitem{ras4}  E. D'Hoker, S.D. Mathur, A. Matusis and L. Rastelli, Nucl.Phys. {\bf B589} (2000) 38.

\bibitem{sanjay}  Sanjay, Mod.Phys.Lett. {\bf A14} (1999) 1413.

\bibitem{mo1}  J.M. Maldacena and Y. Ooguri, J.Math.Phys. {\bf42} (2001) 2929.

\bibitem{kutasov} A. Giveon and D. Kutasov, JHEP {\bf 0001} (2000) 023.

\bibitem{ooguri}  J. de Boer, H. Ooguri, H. Robins and J. Tannenhauser, JHEP {\bf 12} (1998) 026.

\bibitem{fz4p}   A.B. Zamolodchikov and V.A. Fateev, Sov.J.Nucl.Phys. {\bf 43} (4) (1986) 657.

\bibitem{gastonlog} G. Giribet, Mod.Phys.Lett. {\bf A16} (2001) 821. See version arXiv:hep-th/0105248v3

\bibitem{becker}  K. Becker and M. Becker, Nucl.Phys. {\bf B418} (1994) 206.

\bibitem{teschnerliouville} J. Teschner, Class.Quant.Grav. 18 (2001) R153. J. McGreevy, J. Teschner and H. Verlinde, {\it Classical and Quantum D-branes in 2D String Theory}, arXiv: hep-th/0305194.

\bibitem{zamo2} V. Fateev, A. Zamolodchikov and Al. Zamolodchikov, {\it Boundary Liouville Field Theory I. Boundary State and Boundary Two-point Function}, arXiv:hep-th/0001012.

\bibitem{nos3}  G. Giribet and C. N\'{u}\~{n}ez, JHEP {\bf 0106} (2001) 010.

\bibitem{nichols}  A. Nichols and Sanjay, Nucl.Phys. {\bf B597} (2001) 633-651

\bibitem{a4p}  O. Andreev, Phys.Lett. {\bf B363} (1995) 166.

\bibitem{ay}  H. Awata and Y. Yamada, Mod.Phys.Lett. {\bf A7} 3 (1992) 1185.

\bibitem{petko2}  P. Furlan, A. Ganchev, R. Paunov and V. Petkova, Nucl. Phys. {\bf B394} (1993) 665.

\bibitem{seibergs} Nathan Seiberg and David Shih, {\it Branes, Rings and Matrix Models in Minimal (Super)string Theory}, arXiv:hep-th/0312170. I. R. Klebanov, J. Maldacena and N. Seiberg, {\it Unitary and Complex Matrix Models as 1-d Type 0 Strings}, arXiv:hep-th/0309168. M. R. Douglas, I. R. Klebanov, D. Kutasov, J. Maldacena, E. Martinec and N. Seiberg {\it A New Hat For The c=1 Matrix Model}, arXiv:hep-th/0307195. Igor R. Klebanov, Juan Maldacena and Nathan Seiberg, {\it  D-brane Decay in Two-Dimensional String Theory}, JHEP {\bf 0307} (2003) 045. Jan de Boer, Annamaria Sinkovics, Erik Verlinde and Jung-Tay Yee, {\it String Interactions in c=1 Matrix Model}, arXiv:hep-th/0312135. John McGreevy, Sameer Murthy and Herman Verlinde, {\it Two-dimensional superstrings and the supersymmetric matrix model}, arXiv:hep-th/0308105. John McGreevy and Herman Verlinde, JHEP {\bf 0312} (2003) 054. Yu Nakayama, {\it Liouville Field Theory - A decade after the revolution}, arXiv:hep-th/0402009.

\bibitem{teschnerotra} J. Teschner, Phys. Lett. {\bf B363} (1995) 65. A. Zamolodchikov and Al. Zamolodchikov, Nucl.Phys. {\bf B477} (1996) 577-605.

\bibitem{teschnernuevos}  J. Teschner, {\it Quantum Liouville theory versus quantized Teichm\"uller spaces}, Fortsch.Phys. {\bf 51} (2003) 865. J. Teschner, {\it On the relation between quantum Liouville theory and the quantized Teichm\"uller spaces}, Contribution to the proceedings of the 6th International Conference on CFTs and Integrable Models, Chernogolovka, Russia, September 2002, arXiv:hep-th/0303149. J. Teschner, {\it  A lecture on the Liouville vertex operators},  Contribution to the proceedings of the 6th International Conference on CFTs and Integrable Models, Chernogolovka, Russia, 2002, arXiv:hep-th/0303150. J. Teschner {\it From Liouville Theory to the Quantum Geometry of Riemann Surfaces},  Contribution to the Proceedings of the ICMP 2003, Lisbon, arXiv:hep-th/0308031.
  

\bibitem{ultimo} Al. Zamolodchikov, {\it Higher equations of motion in Liouville field theory}, arXiv:hep-th/0312279

\bibitem{si} S.i. Yamaguchi, Phys.Lett. {\bf B546} (2002) 300.

\bibitem{mitesisdoc} G. Giribet, {\it String theory on $AdS_3 \times {\mathcal N}$}, Doctoral (PhD) thesis (2003), Departamento de F\'{\i}sica de la Facultad de Ciencias Exactas y Naturales de la Universidad Buenos Aires FCEN-UBA. Bibliography reference: http://www.opac.bl.fcen.uba.ar/index.html Thesis 3586.

\end{thebibliography}
\end{document}